\documentclass[aps,pre,twocolumn,groupedaddress]{revtex4-1}
\usepackage[pdftex]{graphicx}
\usepackage{bm}
\usepackage{amsmath}
\usepackage{color}

\begin{document}

%\preprint{}

\title{First passage time properties of diffusion with a broad class of stochastic diffusion coefficients}

\author{Go Uchida}
\affiliation{Department of Mechanical Systems Engineering, 
Tokyo Metropolitan University, Tokyo 1920397, Japan}
\affiliation{Graduate School of Information Science, 
University of Hyogo, Hyogo 6500047, Japan}
\author{Hiromi Miyoshi}
\affiliation{Department of Mechanical Systems Engineering, 
Tokyo Metropolitan University, Tokyo 1920397, Japan}
\author{Hitoshi Washizu}
\affiliation{Graduate School of Information Science, 
University of Hyogo, Hyogo 6500047, Japan}
%\homepage[]{Your web page
%\thanks{}


\date{\today}

\begin{abstract}
This study investigates the first passage time (FPT) properties of particles with a broad class of positive stochastic diffusion coefficients (DCs), 
representing diffusion in heterogeneous environments or of particles with conformational fluctuations. 
We demonstrate that for diffusion in a one-dimensional semi-infinite domain with an absorbing boundary, 
particles will eventually reach the absorbing boundary with probability one. 
We also show that a stochastic DC provides higher transport efficiency in an early arrival of particles at the boundary 
than would be expected under diffusion whose DC is the ensemble average of the stochastic DC. 
Furthermore, 
a stochastic DC with a larger supremum exhibits a more efficient transport even if ensemble averages are the same. 
For ergodic DCs, 
we show three more properties: the mean FPT diverges, the enhancement of early-arrival efficiency diminishes over long times, 
and the FPT distribution converges to a L{\'e}vy-Smirnov distribution in the long-time limit. 
These properties are shown to arise from the convergence of the time-averaged DC to the ensemble average, 
with the convergence speed determined by the DC's fluctuation time scale. 
We finally discuss the similarities and differences of FPT properties between three-dimensional diffusion outside a spherical absorbing boundary and 
the one-dimensional diffusion. 
%the diffusion in one-dimensional semi-infinite domain with an absorbing boundary.
Our results indicate that fluctuations in DCs may need to be non-Markov and/or non-ergodic to allow efficient transport of particles to distant targets. 
Our results also suggest that 
fluctuations in a DC play an important role, for example, in diffusion-limited reactions triggered by single molecules  
in physics, chemistry, or biology.  
\end{abstract}

%\keywords{}

\maketitle

\section{Introduction}
%\subsection{}
%\subsubsection{}
Fluctuations in diffusion coefficients (DCs) have been reported in various physical and biological systems~\cite{Leith12, Parry14, Yamamoto98, Yamamoto21, Manzo15}. 
For example, 
molecular dynamics simulations of supercooled liquids demonstrate a spatio-temporal heterogeneity of diffusivity, 
which results in a temporal fluctuation of the DC along a trajectory of a particle~\cite{Yamamoto98}. 
Molecular dynamics simulations of small proteins show temporal fluctuations in the DCs due to conformational fluctuations~\cite{Yamamoto21}.  
In addition, 
an experimental observation of 
subdiffusion of a certain receptor on a cell membrane is explained by 
a model of a Brownian motion with a stochastic DC~\cite{Manzo15}. 

Non-Gaussian yet Brownian diffusion, 
where the mean squared displacement grows linearly in time 
but the propagator for displacement is non-Gaussian, 
has been widely observed in experiments involving colloidal particles, biological systems, and soft matter~\cite{Wang09, Wang12, Guan14, Xue16, Alexandre22}. 
Experimental studies reveal that such dynamics often manifest as exponential or stretched exponential propagator at short times, 
with a crossover to Gaussian distributions at longer times,   
reflecting the underlying heterogeneity and temporal fluctuations in the environment~\cite{Wang09, Wang12}.  

To explain the non-Gaussianity of the propagator, 
Granick et al. have proposed the use of Beck and Cohen's superstatistical approach~\cite{Wang12, Beck03, Beck05, Beck06}. 
In this approach, 
the environment is viewed as locally homogeneous but globally heterogeneous, 
where the DC follows a specific probability distribution, 
representing a superposed ensemble of Brownian motions with different DCs. 
This approach leads to the mean squared displacement proportional to time. 
In addition, 
depending on the chosen distribution of the DC (such as exponential distribution, gamma distribution, or power-law form), 
this approach 
%explains the linear time dependence of the mean squared displacement and 
successfully derives the propagators with non-Gaussian features like Laplace or power-law tails~\cite{Chechkin17, Jain17, Jain17b, Hapca09}. 
%While successful in certain regimes, 
However, 
the superstatistical approach cannot describe a crossover of the propagator to a Gaussian distribution over extended timescales, 
because it assumes a time-invariant distribution of the DC. 

The diffusing diffusivity (DD) model has emerged as a central framework for explaining the observation of a non-Gaussian propagator at short times and 
its crossover to a Gaussian distribution at longer times~\cite{Chubynsky14}. 
In this framework, 
the DC itself is treated as a stochastic process to capture the effects of environmental fluctuations and heterogeneity. 
Since the proposal of the model, 
the DD approach incorporating various stochastic process models for the DC and 
its extensions have been extensively investigated ~\cite{Jain16, Jain17, Tyagi17, Chechkin17, Sposini18, Lanoiselee18, Lanoiselee18b, Sposini19, Metzler20, Luo24, Akimoto26}. 
Both analytical and numerical studies, 
specifically using specific stochastic processes for the DC, 
demonstrate that the DD model can reproduce key experimental features, 
such as the persistence of non-Gaussianity at short times and the eventual recovery of Gaussian behavior ~\cite{Chechkin17, Sposini18, Sposini19}.

%%Theoretically, 
%The diffusing diffusivity (DD) model has emerged as a central framework to explain the observations of a non-Gaussian propagator at short times and 
%a crossover of the propagator to a Gaussian distribution at longer times~\cite{Chubynsky14}.  
%%these observations~\cite{Chubynsky14}. 
%In this model, the DC itself is treated as a stochastic process, 
%capturing the effects of environmental fluctuations and heterogeneity. 
%%~\cite{Chechkin17, Sposini18, Lanoiselee18, Sposini19, Luo24}. 
%Following the proposal of the model, 
%further developments and extensions of this approach have been extensively reported~\cite{Jain16, Jain17, Tyagi17, Chechkin17, Sposini18, Lanoiselee18, Lanoiselee18b, Sposini19, Metzler20, Luo24, Akimoto26}. 
%%Subsequent developments of this approach were reported in~\cite{Jain16, Jain17, Tyagi17, Chechkin17, Sposini18, Lanoiselee18, Lanoiselee18b, Sposini19, Metzler20, Luo24, Akimoto26}. 
%Both analytical and numerical studies demonstrate that the DD model can reproduce key experimental features, such as the persistence of non-Gaussianity at short times and 
%the eventual recovery of Gaussian behavior~\cite{Chechkin17, Sposini18, Sposini19}. 

Switching diffusivity (SD) model is another class of diffusivity models~\cite{Bressloff17, Bressloff19, Sungkaworn17, Weron17, Godec17, Yamamoto98, Pastore15, Miyaguchi16, Uneyama15, Miyaguchi19, Uneyama19, Grebenkov21, Soria21, Akimoto26}. 
In the DD model, 
the DC changes continuously. 
In contrast, 
the SD model involves diffusivity that switches randomly between discrete values. 
SD models are frequently used for understanding biological systems~\cite{Bressloff17, Bressloff19, Sungkaworn17, Weron17, Godec17}. 
For example, 
an SD model can explain protein concentration gradients in biological systems like the C. elegans zygote~\cite{Bressloff19}. 
Heterogeneous diffusivity of particles in supercooled liquids can also be explained by 
switching between different diffusive states: fast and slow diffusive states of particles~\cite{Yamamoto98, Pastore15}.  
In a theoretical study inspired by diffusion in supercooled liquids, 
Miyaguchi et al. demonstrated that 
when diffusivity fluctuates between fast and slow states with power-law sojourn time distributions, 
the propagator for displacement is initially non-Gaussian but converges to Gaussian distributions in the long-time limit~\cite{Miyaguchi16}.   

As described above, 
the diffusion in the specific DD and SD models shares two common properties: 
non-Gaussian behavior of the propagator at short times and 
its subsequent convergence to a Gaussian distribution at long times.  
%Both models can generate non-Gaussian propagator for displacement at short times. 
%Both models can produce convergence of non-Gaussian propagator for displacement to Gaussian distributions in the long-time limit. 
Recently, 
it has been shown that for a broad class of models, 
including both the DD model and the SD model, 
the propagator is non-Gaussian, 
especially with heavy tails~\cite{Sposini24e, Uchida24}.    
It has also been revealed that for a broad class of models, 
the ergodicity of a stochastic DC leads to the convergence of non-Gaussian propagator for displacement to Gaussian distributions~\cite{Uchida24}.     

%As we have already described, 
%the diffusion in the DD model and the SD model has two common properties. 
%Both models can generate non-Gaussian propagator for displacement at short times. 
%Both models can produce convergence of non-Gaussian propagator for displacement to Gaussian distributions in the long-time limit. 
%Recently, 
%it has been shown that for a broad class of models, 
%including both the DD model and the SD model, 
%the propagator for displacement is non-Gaussian, 
%especially with heavy tails~\cite{Sposini24e, Uchida24}.    
%It has also been revealed that for a broad class of models, 
%the ergodicity of a stochastic DC leads to the convergence of non-Gaussian propagator for displacement to Gaussian distributions~\cite{Uchida24}.     

%Fluctuations in DCs significantly impact not only the propagator for displacement but also the first passage time (FPT) statistics. 
Not only the statistical properties of displacement 
but also the behavior of the first passage time (FPT) is an important aspect of diffusion.
The FPT is the time required for a diffusing particle to reach a particular target for the first time. 
Thus, 
the FPT statistics often provide important insights into reaction kinetics, search processes, 
and signal initiation~\cite{Szabo80, Condamin07, Monasterio11, Metzler14, Godec16, Godec17, Chou14, Lanoiselee18}.  
%In heterogeneous media, 
%spatiotemporal variations in diffusivity create dynamic disorder that broadens FPT distributions, 
%increasing the likelihood of both short and long reaction trajectories~\cite{Lanoiselee18}.  
%While disorder generally slows average reaction kinetics, 
%its dynamic nature can benefit individual reaction events by enabling faster molecular search processes~\cite{Lanoiselee18}. 

The FPT of the diffusion with fluctuating DCs shares certain properties with the classical Brownian motion,  
%An important property characterizing the classical Brownian motion, 
defined as diffusion with a constant DC. 
An intriguing property characterizing the classical Brownian motion 
in a one-dimensional semi-infinite domain with an absorbing boundary is that 
all particles will eventually reach the absorbing boundary with probability one, 
but the mean FPT diverges. 
It has been shown that 
these properties also hold for the minimal DD model 
where the DC is described by a squared Ornstein-Uhlenbeck process  
or the DD model where the DC follows a geometric Brownian motion~\cite{Sposini19, Grebenkov21}. 
%In addition, 
%it has been revealed that 
%in the minimal DD model, 
%the FPT distribution approaches the L{\'e}vy-Smirnov distribution, 
%which is the FPT distribution for the classical Brownian motion, 
%in the long-time limit~\cite{Sposini19}. 

On the other hand, 
the FPT of the diffusion with fluctuating DCs has a clear advantage over the classical Brownian motion. 
In heterogeneous media, 
spatiotemporal variations in diffusivity create dynamic disorder that broadens FPT distributions, 
increasing the likelihood of both short and long reaction trajectories~\cite{Lanoiselee18}. 
While disorder generally slows average reaction kinetics, 
its dynamic nature can benefit individual reaction events by enabling faster molecular search processes~\cite{Lanoiselee18}. 
Similarly,
although fluctuating DCs may result in longer mean FPTs~\cite{Sposini24e}, 
a previous study on specific models for the DC has demonstrated that they significantly accelerate the extreme FPT dynamics 
in scenarios involving multiple redundant particles~\cite{Sposini24}.

While the propagator for displacement has been well characterized for a broad class of stochastic DCs, 
it remains unknown whether FPT properties such as the above hold for a broad class of stochastic DCs. 
In the present study, 
to clarify how fluctuations in diffusivity affect the timing of important events such as boundary absorption, 
we investigate FPT properties of diffusion with a broad class of stochastic DCs. 
In Sec.~\ref{sec:model}, 
we describe 
a one-dimensional overdamped Langevin equation with 
a stochastic DC. 
Diffusion with a stochastic DC can be described by 
an overdamped Langevin equation with a stochastic DC. 
In Sec.~\ref{sec:FPe}, 
we present an approach to find a formula of the FPT distribution. 
The approach differs from the subordination approach and the approach of solving the Fokker-Planck equations using eigenfunction expansions~\cite{Lanoiselee18,Grebenkov21,Chechkin17,Lanoiselee24}. 
In our approach, 
we derive a diffusion equation corresponding to the overdamped Langevin equation for each sample path of a stochastic DC. 
We then solve this equation to obtain the FPT distribution for each sample path, 
from which we compute the mean distribution over the ensemble of sample paths. 
%The approach is different from subordination~\cite{Lanoiselee18,Grebenkov21}.  
%In our approach, 
%we derive the diffusion equations corresponding to the overdamped Langevin equation with a stochastic DC and 
%then derive a formula of the FPT distribution  
%by solving the diffusion equations. 
In this section, we derive a formula of the FPT distribution in a one-dimensional semi-infinite domain with an absorbing boundary.    
In Sec.~\ref{sec:prop}, 
we reveal FPT properties mainly on the basis of the FPT distribution.  
In Sec.~\ref{sec:discuss}, 
we discuss 
the relation of our results to the results of other studies, 
the generalization of our results to three-dimensional diffusion,   
the relation of our approach to the superstatistical approach, and 
the significance of our results.  
Finally, 
we present our conclusions in Sec.~\ref{sec:conc}.

\section{Model} \label{sec:model}
The one-dimensional overdamped Langevin equation with a stochastic DC 
is given by 
\begin{equation}
    \frac{dx(t)}{dt} = \sqrt{2D(t)} \xi (t),  \label{eq:pifgt1}
\end{equation}
where 
$ x(t) $ is the position of the diffusing particle. 
In Eq.~(\ref{eq:pifgt1}), 
$ D(t) $ is a stochastic process and represents a DC. 
We assume that $ 0 < D(t) $. 
In Eq.~(\ref{eq:pifgt1}), 
$ \xi (t) $ is a Gaussian white noise: $ \langle \xi (t) \rangle = 0 $ and 
$ \langle \xi (t) \xi (t') \rangle = \delta(t-t') $. 
We assume that $ D(t) $ and $ \xi (t) $ are statistically independent.

\section{FPT distribution} \label{sec:FPe}
Here, 
we derive a formula for the FPT distribution. 
There have been two alternative methods for deriving formulas for the FPT distribution: one is the subordination approach and 
the other is solving the Fokker-Planck equation using an eigenfunction expansion~\cite{Chechkin17, Lanoiselee18, Lanoiselee24}. 
Both methods provide a general framework for finding exact solutions of the propagator for displacement and the FPT distribution or 
approximations of them in the short-time limit or the long-time limit for specific stochastic process models for DCs, 
and have played an important role in advancing the theoretical understanding of 
FPT statistics for particles with fluctuating DCs~\cite{Chechkin17, Lanoiselee18, Sposini19, Grebenkov21, Lanoiselee24}.  
However, 
the purpose of the present study is not to clarify the properties of the FPTs  for specific stochastic process models for the DCs, 
but to clarify the properties of the FPTs for a broader class of stochastic DCs.  
Thus, 
we derive a formula using a different method~\cite{Uchida24}  
(see Appendix~\ref{app:equivsofof} for the equivalence of the subordination formula with ours). 
%A formula using subordination has already been derived~\cite{Lanoiselee18},  
%but here we derive a formula using a different method~\cite{Uchida24}
%(see Appendix~\ref{app:equivsofof} for the equivalence of the subordination formula with ours). 

\subsection{Overview of the method}
%\subsection{Derivation of diffusion equations}
The basis of the method is that 
the sample paths of a stochastic DC are not stochastic in time, 
but deterministic.  
In the method, we first derive the diffusion equation for each sample path of a stochastic DC, 
find the propagator for displacement for each sample path by solving the diffusion equation, 
and then obtain the FPT distribution for each sample path from the propagator.
%In the method, 
%we first find the propagator for displacement for each sample path of a stochastic DC, 
%and then obtain the FPT distribution for each sample path from the propagator. 
To derive the FPT distribution of the diffusion described by Eq.~(\ref{eq:pifgt1}), the FPT distribution for each sample path is then averaged over 
the space of the sample paths of the stochastic DC.   
%by averaging the distribution over the space of sample paths of the stochastic DC. 
%The basis of this method is that 
%the sample paths of a stochastic DC are not stochastic in time, 
%but deterministic.  
%the average of the distributions over the space of sample paths of the stochastic DC.
To formalize this approach, 
we begin by defining the diffusion equation for each sample path.

%Here, 
%we derive the diffusion equations corresponding to Eq.~(\ref{eq:pifgt1}). 
%If $ D(t) $ is deterministic and $ \int_{0}^{t} D(t') dt' < \infty $, 
%it is easy to derive the diffusion equation corresponding to Eq.~(\ref{eq:pifgt1})~\cite{Risken}.  
%However, 
%$ D(t) $ is not deterministic by definition.  

%In the present study, 
%we derive the diffusion equations 
%using the method we used in our previous study~\cite{Uchida24}. 
%We firstly derive the diffusion equation for each sample path of a stochastic DC. 
A stochastic process is an ensemble of all possible time functions (sample paths) with a probability defined for subsets of those functions. 
Thus, 
when we denote the sample space for $ D(t) $ as $ \Omega $, 
a sample as $ \omega $, 
and a sample path of $ D(t) $ as $ D(t,\omega ) $,   
Eq.~(\ref{eq:pifgt1}) can be considered as an ensemble of the following equations: 
\begin{equation}
    \frac{dx(t;\omega )}{dt} = \sqrt{2D(t;\omega )} \xi (t) \; (\omega \in \Omega), \label{eq:lefesp}
\end{equation}
where $ x(t;\omega ) $ is the position of the diffusing particle for a given sample path $ D(t,\omega ) $. 
For a given $ \omega $, 
Eq.~(\ref{eq:lefesp}) is the Langevin equation for the sample path of $ D(t;\omega ) $.
Here, 
note that a subset of the equations given by Eq.~(\ref{eq:lefesp}) is assigned the same probability as the corresponding subset of sample paths of $ D(t) $. 
A sample path of a stochastic process is a single, specific outcome from the ensemble of all possible time functions and  
is a deterministic function of time. 
Thus, 
$ D(t,\omega ) $ is a deterministic function of time and 
if $ \int_{0}^{t} D(t';\omega ) dt' < \infty $, 
the diffusion equations corresponding to Eq.~(\ref{eq:lefesp}) is given by 
\begin{equation}
    \partial _{t} G \left(x,t,x_{0};\omega \right) =  D(t;\omega ) \partial _{xx} G \left(x,t,x_{0};\omega \right) \; (\omega \in \Omega), \label{eq:tdfpgt}
\end{equation}
where $ G \left(x,t,x_{0};\omega \right) $ represents the propagator for a given sample path $ D(t;\omega ) $. 
For a given $ \omega $, 
Eq.~(\ref{eq:tdfpgt}) is the diffusion equation for the sample path of $ D(t;\omega ) $.

After solving Eq.~(\ref{eq:tdfpgt}), 
the FPT distribution for each sample path $ g(t,x_{0};\omega ) $ is obtained by evaluating probability flux at an absorbing boundary using $ G \left(x,t,x_{0};\omega \right) $.   
%By averaging $ G \left(x,t,x_{0};\omega \right) $ over $ \Omega $, 
%we also obtain the propagator for the diffusion described by Eq.~(\ref{eq:pifgt1}):  
%\begin{equation}
%    G \left(x,t,x_{0} \right) = \int_{\Omega } G \left(x,t,x_{0};\omega \right) P(d\omega ),    \label{eq:aveGg}
%\end{equation}
%where $ G \left(x,t,x_{0} \right) $ is the propagator for the diffusion described by Eq.~(\ref{eq:pifgt1}) and 
%$ P(\omega ) $ is the probability measure on $ \Omega $. 
%Similarly, 
%In addition, 
The FPT distribution for the diffusion described by Eq.~(\ref{eq:pifgt1}) is then obtained 
by averaging $ g(t,x_{0};\omega ) $ over $ \Omega $, 
%we obtain the FPT distribution for the diffusion described by Eq.~(\ref{eq:pifgt1}):  
\begin{equation}
    g(t,x_{0}) = \int_{\Omega } g(t,x_{0};\omega ) P(d\omega ),    \label{eq:aveGgfptd}
\end{equation}
where $ g(t,x_{0}) $ is the FPT distribution for the diffusion described by Eq.~(\ref{eq:pifgt1}) and 
$ P(\omega ) $ is the probability measure on $ \Omega $.

\subsection{A formula of FPT distribution} \label{sec:nGGf}
Here, 
we solve Eq.~(\ref{eq:tdfpgt}) by using the method of images~\cite{Redner01},  
and derive the FPT distribution for diffusion in a one-dimensional semi-infinite domain with an absorbing boundary. 
We have the initial condition, 
\begin{equation}
    G \left(x,0,x_{0};\omega \right) = \delta (x-x_{0} ),  
\end{equation}
where $ 0 < x_{0} $. 
In addition, 
we have the absorbing boundary condition, 
\begin{equation}
    G \left(0,t,x_{0};\omega \right) = 0.   
\end{equation} 
We also have the natural boundary condition,  
\begin{equation}
    \lim_{x \to \infty }  G \left(x,t,x_{0};\omega \right) = 0. 
\end{equation}
By solving Eq.~(\ref{eq:tdfpgt}) 
under these initial and boundary conditions, 
we have the propagator 
\begin{widetext}
\begin{equation}
    G \left(x,t,x_{0};\omega \right) = \frac{1}{\sqrt{4\pi S(t;\omega ) t}} \left\{ \exp \left[ -\frac{(x-x_{0})^{2}}{4S(t;\omega )t} \right] - \exp \left[ -\frac{(x+x_{0})^{2}}{4S(t;\omega )t} \right]\right\},    \label{eq:Gomega}
\end{equation} 
\end{widetext}
where 
$ S(t;\omega ) $ represents the time average of $ D(t;\omega ) $ 
and 
is given by 
\begin{widetext}
\begin{equation}
    S(t;\omega ) = \frac{1}{t} \int _{0}^{t} D(t';\omega ) dt'. \label{eq:Mgt}
\end{equation}
\end{widetext}
Thus, 
the propagator for the diffusion described by Eq.~(\ref{eq:pifgt1}) 
$ G \left(x,t,x_{0} \right) $ is 
given by 
\begin{widetext}
\begin{equation}
    G \left(x,t,x_{0} \right)  = \int _{0}^{\infty }  \frac{p(S,t)}{\sqrt{4\pi S t}} \left\{ \exp \left[ -\frac{(x-x_{0})^{2}}{4St} \right] - \exp \left[ -\frac{(x+x_{0})^{2}}{4St} \right]\right\} dS, \label{eq:pdexp2gf}
\end{equation}
\end{widetext}
where $ p(S,t) $ represents the probability distribution of $ S(t) $. 

From Eq.~(\ref{eq:Gomega}), 
we have the FPT distribution for a given sample path $ D(t;\omega ) $~\cite{Molini11} 
\begin{equation}
    g(t,x_{0};\omega ) = \frac{x_{0}D(t;\omega )}{\sqrt{4\pi S^{3}(t;\omega ) t^{3}}} \exp \left[ -\frac{x_{0}^{2}}{4S(t;\omega )t} \right]. \label{eq:rhogfgtt}
\end{equation}
%where $ g(t,x_{0};\omega ) $ represents the FPT distribution for a given sample path $ D(t;\omega ) $. 
Thus, 
the FPT distribution for the diffusion described by Eq.~(\ref{eq:pifgt1}) 
$ g(t,x_{0}) $ is 
given by 
\begin{equation}
    g(t,x_{0}) = \int _{\Omega }  \frac{x_{0}D(t;\omega )}{\sqrt{4\pi S^{3}(t;\omega ) t^{3}}} \exp \left[ -\frac{x_{0}^{2}}{4S(t;\omega )t} \right] P(d\omega ). \label{eq:pdexp2gf2}
\end{equation}

\section{FPT properties} \label{sec:prop}
\subsection{General properties} \label{ss:gp}
The particle will eventually reach the absorbing boundary with probability one.
By the variable transformation $ t''(\omega ) = \int_{0}^{t} D(t';\omega ) dt' / x_{0}^{2} $, 
we obtain 
\begin{equation}
    \int_{0}^{\infty } g(t,x_{0};\omega ) dt = 1.  \label{eq:itone}
\end{equation} 
%Note that for Eq.~(\ref{eq:itone}) to hold, 
%it is essential that $ t''(\omega ) \to \infty $ as $ t \to \infty $. 
%In other word, 
%it is required that $ \lim_{t \to \infty } \int_{0}^{t} D(t';\omega ) dt' \to \infty $. 
Thus, 
from Eq.~(\ref{eq:pdexp2gf2}), 
we have 
\begin{equation}
    \int_{0}^{\infty } g(t,x_{0}) dt = 1. 
\end{equation}
This result is consistent with that obtained with the minimal DD model and the DD model where the DC is described by a geometric Brownian motion, 
though the DC of the minimal DD model can be zero~\cite{Sposini19, Grebenkov21}.

Diffusion with a stochastic DC exhibits higher transport efficiency in an early arrival of particles at the absorbing boundary 
than would be expected under diffusion with the deterministic DC of $ \left\langle D(t) \right\rangle $ (the corresponding diffusion with ensemble-averaged diffusivity).
When very few particles with a stochastic DC have yet reach the absorbing boundary, 
from Eq.~(\ref{eq:pdexp2gf}), 
we have 
\begin{equation}
    G \left(x,t,x_{0} \right)  \approx \int _{0}^{\infty }  \frac{p(S,t)}{\sqrt{4\pi S t}} \exp \left[ -\frac{(x-x_{0})^{2}}{4St} \right] dS. \label{eq:pdexp2gf22}
\end{equation} 
Under the same condition, 
the propagator for 
the corresponding diffusion with ensemble-averaged diffusivity is given by 
\begin{equation}
    G_{d} \left(x,t,x_{0} \right)  \approx \frac{1}{\sqrt{4\pi \left\langle S(t) \right\rangle t}} \exp \left[ -\frac{(x-x_{0})^{2}}{4 \left\langle S(t) \right\rangle t} \right] , \label{eq:pdexp2gfd2}
\end{equation} 
where $ G_{d} \left(x,t,x_{0} \right) $ represents the propagator for the corresponding diffusion with ensemble-averaged diffusivity. 
In a previous study, 
we showed that the propagator given by Eq.~(\ref{eq:pdexp2gf22}) has heavier tails than the propagator given by Eq.~(\ref{eq:pdexp2gfd2})~\cite{Uchida24}.  
This means that the short-time side tail of the FPT distribution given by Eq.~(\ref{eq:pdexp2gf2}) is heavier than that of 
the FPT distribution for the corresponding diffusion with ensemble-averaged diffusivity: 
 \begin{equation}
    g_{d} (t,x_{0}) = \frac{x_{0}\left\langle D(t) \right\rangle }{\sqrt{4\pi \left\langle S(t) \right\rangle ^{3} t^{3}}} \exp \left( -\frac{x_{0}^{2}}{4\left\langle S(t) \right\rangle t} \right) .  \label{eq:fptdd}
\end{equation}
Thus, 
the cumulative probability of a particle arriving early at the absorbing boundary for diffusion with a stochastic DC exceeds 
that of the corresponding diffusion with ensemble-averaged diffusivity. 

Next, 
we derive an expression for the excess cumulative probability of early-arriving particles. 
By the variable transformation $ t''(\omega ) = \int_{0}^{t} D(t';\omega ) dt' / x_{0}^{2} $, 
we obtain the cumulative probability of particles arrived at the boundary by time $ t $: 
\begin{equation}
    \int_{0}^{t } g(t',x_{0} ) dt' = \int_{0}^{\infty } p(S,t) \mathrm{erfc} \left( \frac{x_{0}}{2\sqrt{St}} \right) dS, \label{eq:gppp}
\end{equation}
where 
$ \mathrm{erfc} (\cdot ) $ is the complementary error function. 
Similarly, 
from Eq.~(\ref{eq:fptdd}), 
by the variable transformation $ t'' = \int_{0}^{t} \left\langle D(t') \right\rangle dt' / x_{0}^{2} $, 
for the corresponding diffusion with ensemble-averaged diffusivity, 
we obtain the cumulative probability of particles arrived at the boundary by time $ t $: 
\begin{equation}
    \int_{0}^{t} g_{d} (t',x_{0}) dt' =  \mathrm{erfc} \left( \frac{x_{0}}{2\sqrt{ \left\langle S(t) \right\rangle t}} \right) . \label{eq:ppcdd}
\end{equation}
The complementary error function is approximately zero when its argument is larger than or equal to two. 
On the other hand, 
when the argument becomes smaller than two, 
the function becomes larger rapidly. 
Here, 
we denote the solution of the equation $ 4 \sqrt{ \left\langle S(t) \right\rangle t} = x_{0} $ as $ t_{s} $. 
From Eq.~(\ref{eq:ppcdd}) and the equation for $ t_{s} $, 
we can see that $ t_{s} $ represents the time at which 
the cumulative probability of arriving particles in the corresponding diffusion with ensemble-averaged diffusivity begins to rise rapidly from almost zero. 
Thus, 
taking the cumulative probability of particles arrived at the boundary by time $ t_{s} $ in 
the corresponding diffusion with ensemble-averaged diffusivity as a criterion, 
from Eq.~(\ref{eq:gppp}), 
the excess cumulative probability of early-arriving particles is given by 
\begin{widetext}
\begin{equation}
    \int_{0}^{t_{s}} g (t,x_{0}) - g_{d} (t,x_{0}) dt = \int_{0}^{\infty } p(S,t_{s}) \left[ \mathrm{erfc} \left( \frac{2}{\sqrt{S/\left\langle S(t_{s}) \right\rangle}} \right) - \mathrm{erfc} (2) \right] dS .  \label{eq:prg}
\end{equation}
\end{widetext}
From this equation, 
we can see that for a given $ x_{0} $, 
the excess cumulative probability of early-arriving particles is determined only by the time average of $ D(t;\omega ) $.
 
For $ t_{s} $ sufficiently smaller than the time that characterizes the change in $ D(t) $, 
we can show that even if the ensemble averages of stochastic DCs are the same, 
diffusion with a stochastic DC with a larger supremum exhibits a more efficient transport in an early arrival of particles at the absorbing boundary.
When $ t_{s} $ is sufficiently smaller than the time that characterizes the change in $ D(t) $, 
particles begin to reach the absorbing boundary before the time change in $ D(t) $ occurs. 
We have an approximation for the FPT distribution $ g(t,x_{0}) $ on the short-time side, 
\begin{eqnarray}
    g(t,x_{0}) &\approx & \int _{\Omega }  \frac{x_{0}}{\sqrt{4\pi D(0;\omega ) t^{3}}} \exp \left[ -\frac{x_{0}^{2}}{4D(0;\omega )t} \right] P(d\omega ) \nonumber \\
                    &=& \int_{0}^{\infty }  \frac{p_{D}(D,0)x_{0}}{\sqrt{4\pi Dt^{3}}} \exp \left( -\frac{x_{0}^{2}}{4Dt} \right) dD,    \label{eq:fptdstapprox} 
\end{eqnarray}
where $ p_{D}(D,0) $ represents the initial probability distribution of the DC. 
This is a superstatistical approximation of the FPT distribution (see Sec.~\ref{sec:discuss} for details). 
For a DC is described by the Cox-Ingersoll-Ross process, 
superstatistical approximation for the FPT distribution has previously been derived 
for the formula based on the eigenfunction expansion of the Fokker-Planck equation~\cite{Lanoiselee18, Lanoiselee24}. 
In contrast, 
our formula offers a distinct perspective, 
providing an alternative insight into the superstatistical approximation for the FPT distribution.
From Eq.~(\ref{eq:fptdstapprox}), 
we can see that $ g(t,x_{0}) $ on the short-time side is approximated by a superposition of L{\' e}vy-Smirnov distributions. 
Thus,    
even if the ensemble averages of stochastic DCs are the same, 
a stochastic DC with a larger supremum leads to a higher excess cumulative probability of early-arriving particles.

\subsection{Properties for ergodic DCs}    \label{ss:propsfeDC}
In a previous study, 
we showed that when a stochastic DC is ergodic, 
the free propagator crosses over a Gaussian distribution for the corresponding diffusion with ensemble-averaged diffusivity in the long-time limit~\cite{Uchida24}. 
This indicates that when a stochastic DC is ergodic, 
FPT properties also become similar to those of the corresponding diffusion with ensemble-averaged diffusivity in the long-time limit.  

Here, 
we assume that $ D(t) $ is stationary and ergodic. 
Since $ D(t) $ is stationary, 
the ensemble average and the variance of $ D(t) $ are time-independent: 
$ \left\langle D(t) \right\rangle = D_{m} $ and  
$ \left\langle \left( D(t) - D_{m} \right) ^{2} \right\rangle = \sigma ^{2} $. 
We also have $ \lim_{t \to \infty } S\left( t;\omega \right) = D_{m} $ 
because $ D(t) $ is ergodic. 

For later convenience, 
we rewrite $ D(t) $ as $ D(t) = D_{m} h(t) $: $ \langle h(t) \rangle = 1 $, 
$ \left\langle \left( h(t) - 1 \right) ^{2} \right\rangle = \left( \frac{\sigma }{D_{m}} \right) ^{2} $.  
We also define $ H(t) $ as 
$ H(t;\omega ) = \frac{1}{t} \int_{0}^{t} h(t';\omega ) dt' $: $ \lim_{t \to \infty } H(t;\omega ) = 1 $. 

Using 
$ h(t) $ and $ H(t) $, 
we can rewrite Eqs.~(\ref{eq:pdexp2gf2}) and (\ref{eq:fptdd}) as 
\begin{eqnarray}
    g(t,x_{0}) &=& \int_{\Omega } \frac{x_{0} h(t;\omega )}{\sqrt{4\pi D_{m} H^{3}(t;\omega ) t^{3}}} \nonumber \\
                &  & \times \exp \left[ -\frac{x_{0}^{2}}{4D_{m} H(t;\omega )t} \right] P(d\omega ), \label{eq:rhogfgttdl} \\
    g_{d} (t,x_{0}) &=& \frac{x_{0}}{\sqrt{4\pi D_{m} t^{3}}} \exp \left( -\frac{x_{0}^{2}}{4D_{m}t} \right) .  \label{eq:gbm2}
\end{eqnarray}
From Eq.~(\ref{eq:gbm2}), 
we can see that the FPT distribution for the corresponding diffusion with ensemble-averaged diffusivity is 
the L{\' e}vy-Smirnov distribution.  
We can also rewrite Eqs.~(\ref{eq:gppp}) and (\ref{eq:ppcdd}) as 
\begin{eqnarray}
    \int_{0}^{t} g(t',x_{0}) dt' &=& \int_{0}^{\infty } q(H,t) \nonumber \\
                                    & & \times \mathrm{erfc} \left( \frac{x_{0}}{2\sqrt{D_{m}Ht}} \right) dH,  \label{eq:ppsdc} \\
    \int_{0}^{t} g_{d} (t',x_{0}) dt' &=& \mathrm{erfc} \left( \frac{x_{0}}{2\sqrt{D_{m}t}} \right) ,   \label{eq:propgbm}
\end{eqnarray}
where $ q(H,t) $ represents the probability distribution of $ H(t) $. 

For both diffusion with an ergodic DC and the corresponding diffusion with ensemble-averaged diffusivity, 
the mean FPT is infinite. 
For a large $ t $, 
from Eq.~(\ref{eq:gbm2}), 
we have 
\begin{equation}
    g_{d} (t,x_{0}) \sim \frac{x_{0}}{\sqrt{4\pi D_{m} t^{3}}} .   \label{eq:gsDd}
\end{equation} 
Thus, 
the mean FPT is infinite. 
Similarly, 
for a large $ t $, 
from Eq.~(\ref{eq:rhogfgttdl}), 
we have 
\begin{equation}
    g(t,x_{0}) \sim \frac{x_{0}}{\sqrt{4\pi D_{m} t^{3}}} ,   \label{eq:gsD}
\end{equation} 
because $ \lim_{t \to \infty } H(t;\omega ) = 1 $ and $ \langle h(t) \rangle = 1 $. 
Thus, 
the mean FPT is infinite. 
This result is consistent with those obtained with the minimal DD model and 
the DD model described by the Cox-Ingersoll-Ross process~\cite{Sposini19, Lanoiselee18}.  

From Eq.~(\ref{eq:ppsdc}), 
for ergodic DCs,  
we have the excess cumulative probability of early-arriving particles, 
\begin{widetext}
\begin{equation}
    \int_{0}^{t_{s}} g(t,x_{0}) - g_{d} (t,x_{0}) dt = \int_{0}^{\infty } q(H,t_{s} ) \left[ \mathrm{erfc} \left( \frac{2}{\sqrt{H}} \right) - \mathrm{erfc} (2) \right] dH,   \label{eq:pr}
\end{equation} 
\end{widetext}
where $ t_{s} = \frac{x_{0}^{2}}{16D_{m} } $. 
From Eq.~(\ref{eq:pr}), 
when $ t_{s} \to \infty $, 
the excess cumulative probability of early-arriving particles approaches zero 
because 
\begin{equation}
    \lim_{t \to \infty } q(H,t ) = \delta (H -1) .    \label{eq:limH}
\end{equation} 

In fact, 
when $ t_{s} \rightarrow \infty $, 
the FPT distribution for diffusion with an ergodic DC approaches the FPT distribution 
for the corresponding diffusion with ensemble-averaged diffusivity. 
From Eqs.~(\ref{eq:rhogfgttdl}) and (\ref{eq:limH}), 
when $ t_{s} \rightarrow \infty $, 
we have 
\begin{equation}
    g (t,x_{0}) \sim \frac{x_{0}}{\sqrt{4\pi D_{m} t^{3}}} \exp \left( -\frac{x_{0}^{2}}{4D_{m}t} \right) .  \label{eq:gtiapprox}
\end{equation}
From Eqs.~(\ref{eq:gbm2}) and (\ref{eq:gtiapprox}), 
we can see that the right-hand side of Eq. (\ref{eq:gtiapprox}) is the FPT distribution for the corresponding diffusion with ensemble-averaged diffusivity.
This result is consistent with that obtained with the minimal DD model~\cite{Sposini19}.  

As we have just shown, 
the convergence of the excess cumulative probability of early-arriving particles to zero and 
the asymptotic approach of the FPT distribution to the L{\' e}vy-Smirnov distribution 
result from the convergence of $ H(t) $ to one. 
Thus, 
the speed of convergence of the excess cumulative probability of early-arriving particles to zero and 
the speed of approach of the FPT distribution to the L{\' e}vy-Smirnov distribution are 
determined by the speed of convergence of $ H(t) $ to one.    
The speed of convergence of $ H(t) $ to one is determined by 
the autocovariance function of $ h(t) $. 
From Eq.~(\ref{eq:Mgt}), 
we have 
\begin{equation}
    \left\langle \left( H\left( t \right) - 1 \right) ^{2}\right\rangle = \frac{1}{t^{2}} \int_{0}^{t} \int_{0}^{t} C(t',t'') dt'dt'',    \label{eq:SvarDcorr}
\end{equation}
where $ C(t',t'') $ represents the autocovariance function of $ h(t) $ and is given by 
\begin{equation}
    C(t',t'') = \left\langle \delta h(t') \delta h(t'') \right\rangle . \label{eq:krtg}
\end{equation}
In this equation, 
$ \delta h(t) $ is given by 
$ \delta h(t) = h(t) - 1 $. 
The autocovariance function depends only on the time difference, 
because $ h(t) $ is stationary. 
Thus, 
we have 
\begin{equation} 
    \left\langle \left( H\left( t \right) - 1 \right) ^{2}\right\rangle = \frac{2}{t^{2}} \int_{0}^{t} \left( t-t' \right) C(t') dt'.    \label{eq:SvarDcorrep}
\end{equation}

For an ergodic DC, 
when $ H(t) $ approaches the neighborhood of one in a finite time, 
we can estimate the limiting distance over which early-arriving particles can be efficiently transported. 
Here, 
we denote the time at which $ H(t) $ reaches the neighborhood of one as $ t_{0} $. 
As we have already shown, 
when $ t_{0} \leq t_{s} $, 
the excess cumulative probability of early-arriving particles is almost zero. 
Thus, 
the limiting distance is given by 
\begin{equation}
    16D_{m} t_{0} > x_{0}^{2} .     \label{eq:cepp}
\end{equation}
Here, 
note that $ t_{s} = \frac{x_{0}^{2}}{16D_{m}} $. 
From Eqs.~(\ref{eq:SvarDcorrep}) and (\ref{eq:cepp}), 
the limiting distance is determined only by 
the ensemble average and the autocovariance function of the DC, 
and does not depend on higher-order correlations of the DC.

\subsection{Case study} \label{ss:cs}
Here, 
we assume that $ D(t) $ is described by a two-state Markov process. 
We also assume that the initial distribution of the DC is the equilibrium distribution. 
Thus, 
the process is stationary. 
We label one state $ + $ and the other state $ - $. 
The DC is equal to $ D_{m} h_{+} $ at $ + $ state and 
is equal to $ D_{m} h_{-} $ at $ - $ state. 
Here, 
$ h_{+} $ and $ h_{-} $ are positive constants and $ h_{-} < h_{+} $.
For simplicity, 
we set both the transition probability from $ + $ state to $ - $ state and 
the transition probability from $ - $ state to $ + $ state to be the same.
The distribution of sojourn time in each state is given by 
\begin{equation}
    \psi (\tau ) = \lambda  e^{-\lambda \tau },    \label{eq:edp} 
\end{equation}
where $ \psi (\tau ) $ is  
the distribution of sojourn time in $ + $ state and $ - $ state, 
and $ \lambda $ is the transition probability. 
We set the value of $ \lambda $ to 0.5. 
In addition, 
unless otherwise specified, 
we set $ D_{m} $ to 0.6, 
$ h_{+} $ to 5/3, 
and $ h_{-} $ to $ 1/3 $. 

For this model, 
we have 
\begin{equation}
    \sigma ^{2} = \left[ \frac{ D_{m} \left( h_{+} - h_{-} \right) }{2} \right] ^{2} . 
\end{equation}
For the model, 
we have the autocovariance function $ C(\Delta t) $: 
\begin{equation}
    C(\Delta t) = \sigma ^{2} e^{-\frac{|\Delta t|}{t_{c}}},    \label{eq:tmacov}
\end{equation}
where $ \Delta t $ is the time difference, 
and $ t_{c} $ is the time constant and is given by $ t_{c} = 1/(2\lambda ) $. 

Substituting Eq.~(\ref{eq:tmacov}) into Eq.~(\ref{eq:SvarDcorrep}) leads to 
\begin{equation}
    \left\langle \left( H\left( t \right) - 1 \right) ^{2}\right\rangle = \frac{ \left( h_{+} - h_{-} \right)^{2} t_{c}^{2} \left( \frac{t}{t_{c}}+e^{-\frac{t}{t_{c}}}-1 \right)}{2{t}^{2}}.    \label{eq:Hvar}
\end{equation}
By putting $ t_{l} = t/t_{c} $ in this equation, 
we have
\begin{equation}
    \left\langle \left( H\left( t_{l} \right) - 1 \right) ^{2}\right\rangle = \frac{ \left( h_{+} - h_{-} \right)^{2} \left( t_{l}+e^{-t_{l}}-1 \right)}{2{t}_{l}^{2}}.    \label{eq:Hvarn}
\end{equation}
\begin{figure}
\includegraphics[width=8.6cm]{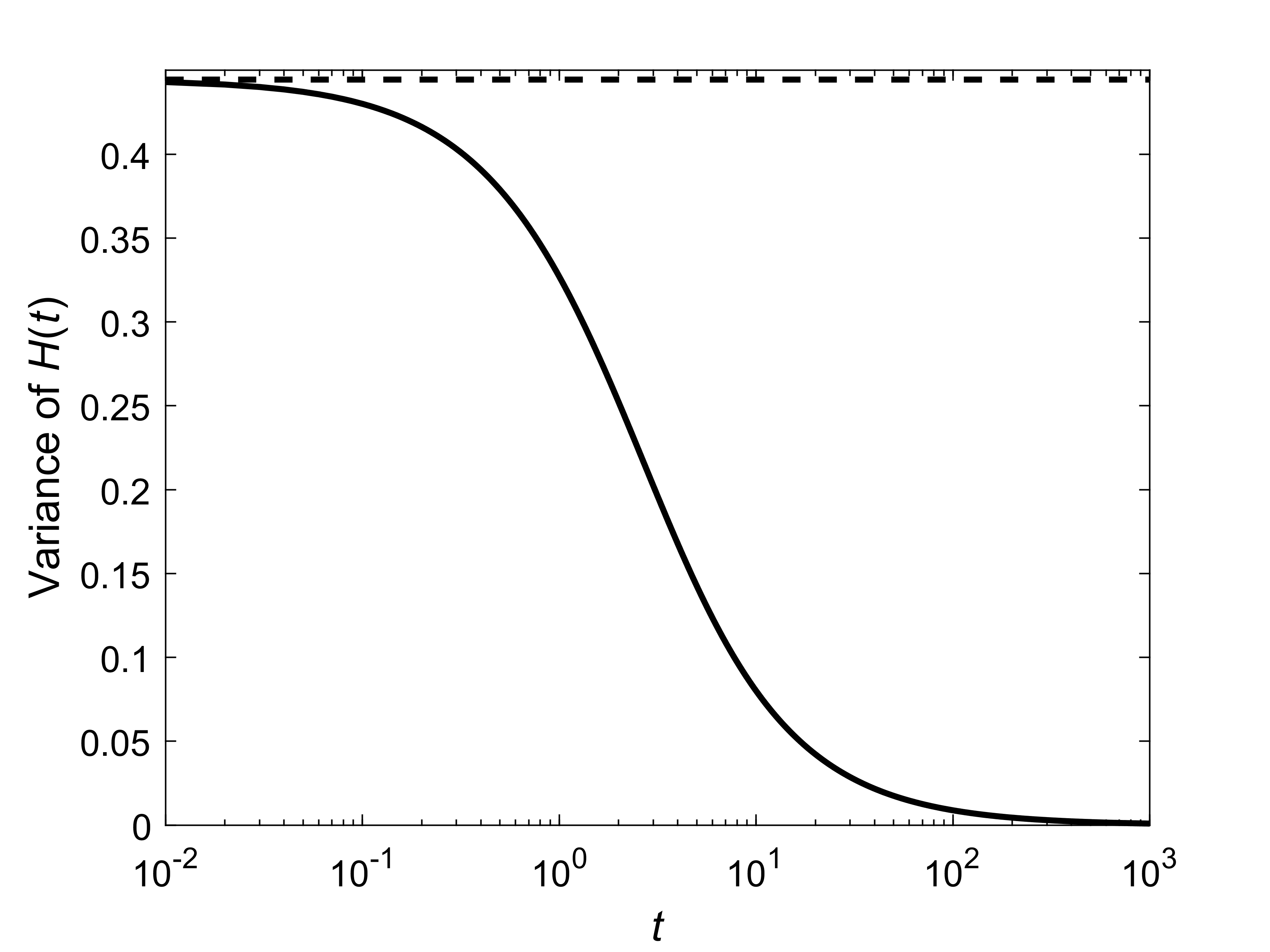}%
\caption{\label{fig:tsH} Time dependence of the variance of $ H(t) $. 
The solid line shows the time dependence estimated from Eq.~(\ref{eq:Hvarn}). 
The dashed line shows the value for short times. 
The value is given by $ \left( h_{+} - h_{-} \right)^{2} / 4 $. 
Time is normalized by $ t_{c} $. 
}
\end{figure}
Figure~\ref{fig:tsH} shows the time dependence of $ \left\langle \left( H\left( t_{l} \right) - 1 \right) ^{2}\right\rangle $. 
We can see that for short times, 
the value of $ \left\langle \left( H\left( t_{l} \right) - 1 \right) ^{2}\right\rangle $ is nearly equal to %$ \sigma ^{2} /D_{m}^{2} $, 
$ \left( h_{+} - h_{-} \right)^{2} / 4 $, 
but rapidly decreases to around zero from time one to time 10. 

Here, 
following the research of Uneyama et al.~\cite{Uneyama15}, 
we find the crossover time of the variance of $ H(t) $. 
We have the asymptotic form for $ t \ll t_{c} $:  
\begin{equation}
    \left\langle \left( H\left( t \right) - 1 \right) ^{2}\right\rangle \approx \frac{ \left( h_{+} - h_{-} \right)^{2}}{4}.    \label{eq:Hvarapprox1}
\end{equation}
We also have the asymptotic form for $ t \gg t_{c} $:  
\begin{equation}
    \left\langle \left( H\left( t \right) - 1 \right) ^{2}\right\rangle \approx \frac{ \left( h_{+} - h_{-} \right)^{2}t_{c}}{2t}.    \label{eq:Hvarapprox2}
\end{equation}
From Eqs.~(\ref{eq:Hvarapprox1}) and (\ref{eq:Hvarapprox2}), 
the crossover time $ t_{cr} $ is estimated as $ t_{cr} = 2t_{c} $. 
This result is consistent with the crossover time of the relative standard deviation (RSD) of the time-averaged squared displacement (TAMSD), 
which is known to reflect the relaxation time of the underlying diffusivity fluctuations~\cite{Uneyama15}. 
By normalizing by $ t_{c} $, 
we have $ t_{lcr} = 2 $. 
Here, $ t_{lcr} = t_{cr} / t_{c} $. 
The value of the crossover time $ t_{lcr} $ is consistent with the observation that 
for short times, 
the value of $ \left\langle \left( H\left( t_{l} \right) - 1 \right) ^{2}\right\rangle $ is nearly equal to %$ \sigma ^{2} /D_{m}^{2} $, 
$ \left( h_{+} - h_{-} \right)^{2} / 4 $, 
but rapidly decreases to around zero from time one to time 10. 

\begin{figure*}
\includegraphics[width=17.8cm]{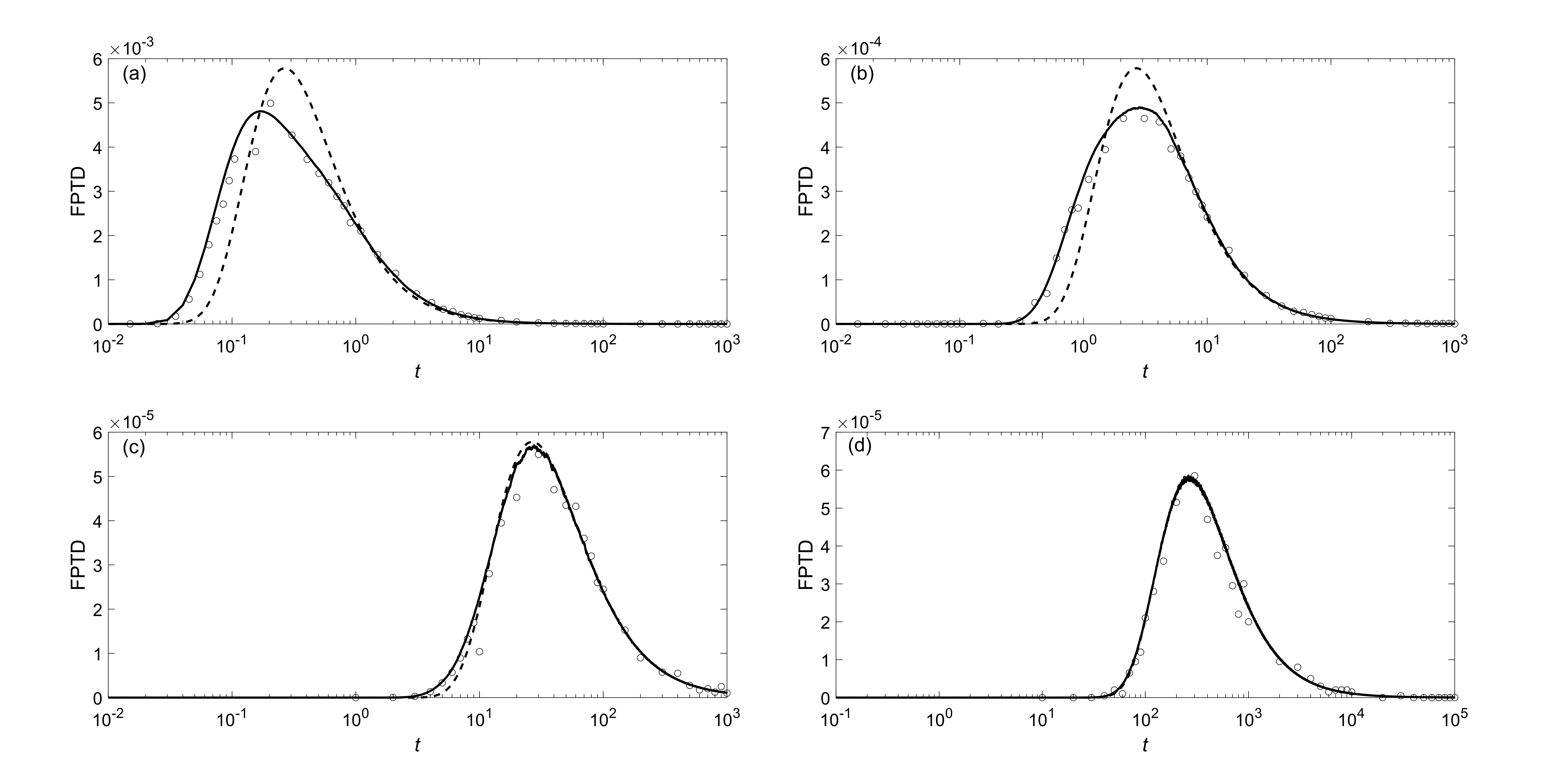}%
\caption{\label{fig:mfptd} FPT distributions (FPTDs) at different values of $ t_{s} $. 
The solid lines show the distributions estimated from Eq.~(\ref{eq:rhogfgttdl}). 
The open circles show the distributions estimated from simulations.  
The dashed lines show the distributions for the corresponding diffusion with ensemble-averaged diffusivity and estimated from Eq.~(\ref{eq:gbm2}). 
(a) $ t_{s} = 0.1 $. (b) $ t_{s} = 1 $. (c) $ t_{s} = 10 $. (d) $ t_{s} = 100 $. 
}
\end{figure*}
Figure~\ref{fig:mfptd} shows the FPT distributions at different values of $ t_{s} $.  
We can see that the distribution estimated from Eq.~(\ref{eq:rhogfgttdl}) is in good agreement with that estimated from simulations 
(see Appendix~\ref{app:sim} for simulation details).  
In addition, 
regardless of $ t_{s} $, 
the distributions for the model in the region $ 100 < t $ closely match the distributions for 
the corresponding diffusion with ensemble-averaged diffusivity.  
This agreement is explained by the variance of $ H(t) $ becoming almost zero at $ t = 100 $.  
In addition, 
from Fig.~\ref{fig:mfptd}(d),  
we can see that when $ t_{s} = 100 $, 
the distribution estimated from Eq.~(\ref{eq:rhogfgttdl}) closely matches the distribution for the corresponding diffusion with ensemble-averaged diffusivity. 
This result is consistent with the theoretical prediction in Sec.~\ref{ss:propsfeDC}:  
as $ H(t) $ approaches one, 
the FPT distribution approaches the L{\' e}vy-Smirnov distribution. 
From Fig.~\ref{fig:mfptd}, 
we can also see that 
the FPT distribution approaches the L{\' e}vy-Smirnov distribution slowly 
when $ t_{s} < t_{lcr} $ and rapidly when $ t_{s} > t_{lcr} $.  

\begin{figure}
\includegraphics[width=8.6cm]{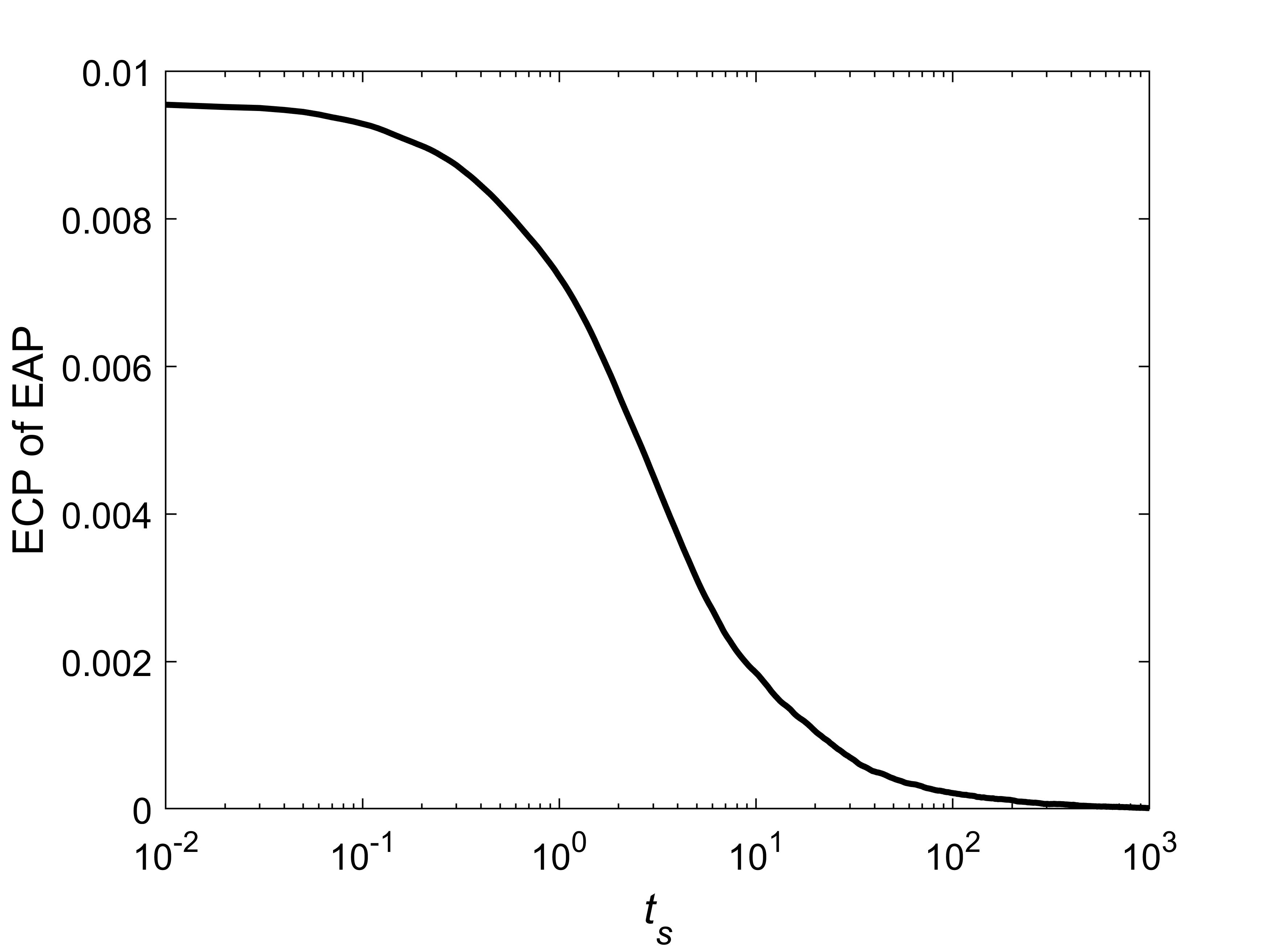}%
\caption{\label{fig:tsppp} The dependence of the excess cumulative probability (ECP) of early-arriving particles (EAP) on $ t_{s} $. 
The solid line shows the dependence estimated from Eq.~(\ref{eq:pr}). 
}
\end{figure}
Figure~\ref{fig:tsppp} shows the dependence of the excess cumulative probability of early-arriving particles  
on $ t_{s} $.  
We can see that the excess cumulative probability of early-arriving particles is nearly equal to $ 0.0095 $ for a small $ t_{s} $ and  
rapidly decreases to around zero from time one to time 10.
This result is consistent with the theoretical prediction in Sec.~\ref{ss:propsfeDC}:  
as $ H(t) $ approaches one, 
the excess cumulative probability of early-arriving particles approaches zero. 
In addition, 
the result indicates that 
fluctuations of the DC are advantageous for efficient transport as long as $ t_{s} < t_{lcr} $,  
but this advantage is rapidly lost as $ t_{s} > t_{lcr} $ and $ t_{s} $ becomes larger. 

Here, 
we estimate the limiting distance. 
From Fig.~\ref{fig:tsH}, 
we can see that we can set $ t_{0} $ to 100. 
Thus, 
from Eq.~(\ref{eq:cepp}), 
we can see that the limiting distance is about 31. 

\begin{figure}
\includegraphics[width=8.6cm]{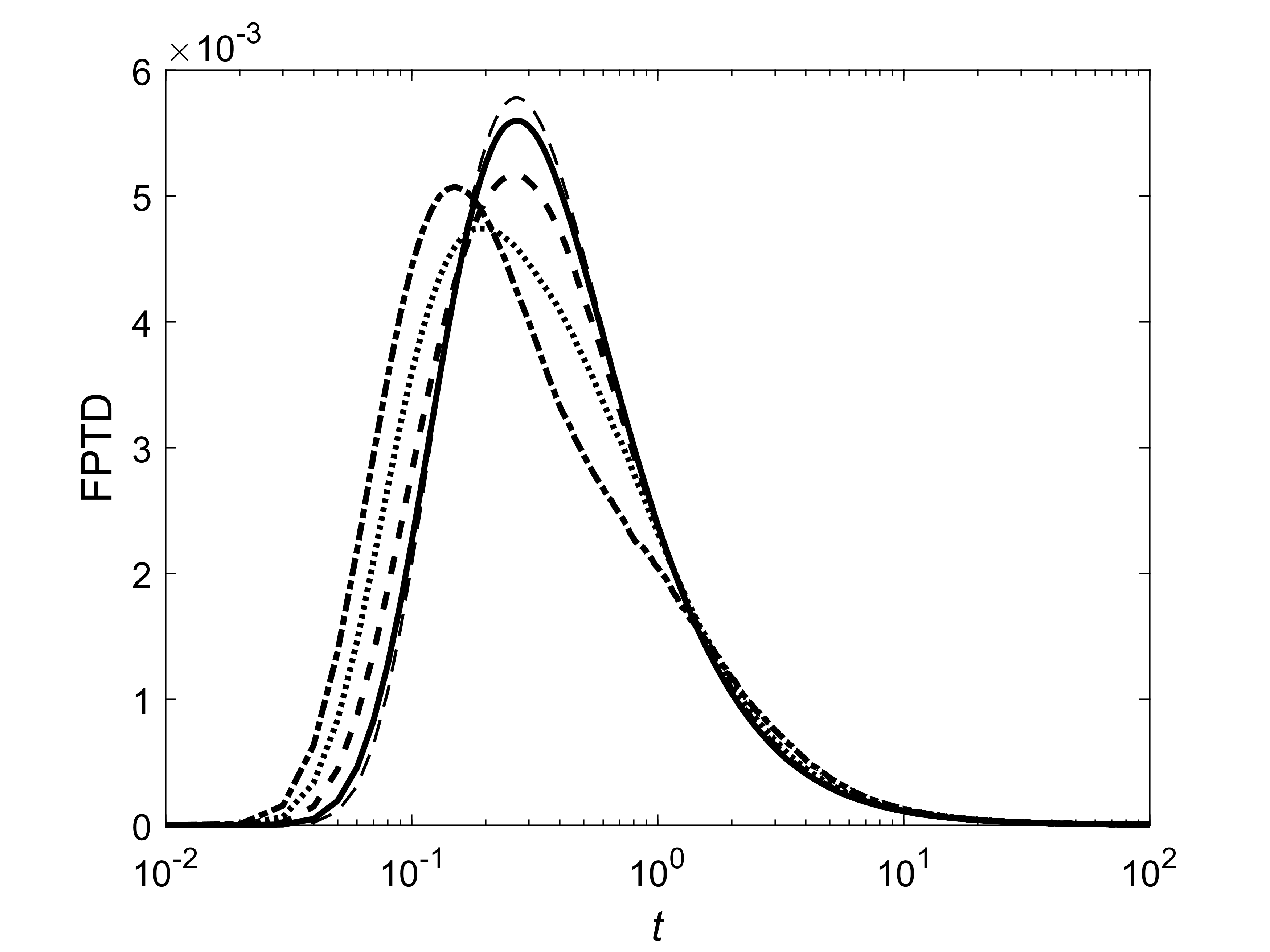}%
\caption{\label{fig:mfdc} Comparison of the FPT distributions for different combinations of $ h_{+} $ and $ h_{-} $ values. 
The solid line shows the distribution for $ h_{+} = 1.2 $ and $ h_{-} = 0.8 $.  
The thick dashed line shows the distribution for $ h_{+} = 1.4 $ and $ h_{-} = 0.6 $. 
The dotted line shows the distribution for $ h_{+} = 1.6 $ and $ h_{-} = 0.4 $. 
The dash-dotted line shows the distribution for $ h_{+} = 1.8 $ and $ h_{-} = 0.2 $. 
The distributions are estimated from Eq.~(\ref{eq:rhogfgttdl}). 
The thin dashed line shows the distribution for the corresponding diffusion with ensemble-averaged diffusivity and estimated from Eq.~(\ref{eq:gbm2}). 
$ t_{s} = 0.1 $ and $ D_{m} = 0.5 $. 
}
\end{figure}
Figure~\ref{fig:mfdc} shows 
comparison of the FPT distributions for different combinations of $ h_{+} $ and $ h_{-} $ values. 
The distributions are for $ t_{s} $ smaller than $ t_{c} $. 
We can see that regardless of the combination of $ h_{+} $ and $ h_{-} $ values, 
the time at which the FPT distribution of the model starts to rise is earlier 
than that of the corresponding diffusion with ensemble-averaged diffusivity. 
We can also see that although the average values of $ h_{+} $ and $ h_{-} $ are the same, 
the larger $ h_{+} $ is, 
the earlier the distribution starts to rise. 
This result is consistent with the theoretical prediction in Sec.~\ref{ss:gp}:  
even though the ensemble averages of the stochastic DCs are the same, 
diffusion with a stochastic DC with a larger supremum exhibits a more efficient transport in an early arrival of particles at the absorbing boundary.

\section{Discussion} \label{sec:discuss}
In the present study, 
we investigated FPT properties of diffusion with a broad class of stochastic DCs that are positive and non-zero. 
We showed that 
for diffusion in a one-dimensional semi-infinite domain with an absorbing boundary, particles will eventually reach the absorbing boundary with probability one. 
We also showed that 
diffusion with a stochastic DC displays higher transport efficiency in an early arrival of particles at the absorbing boundary than would be expected under the corresponding diffusion with ensemble-averaged diffusivity.
For a given distance to the absorbing boundary, 
the excess cumulative probability of early-arriving particles is determined by the probability distribution of the time average of a stochastic DC. 
Moreover, 
when particles begin to reach the absorbing boundary before the time change in a stochastic DC occurs, 
a stochastic DC with a larger supremum shows a more efficient transport in an early arrival of particles at the absorbing boundary 
even if the ensemble averages of stochastic DCs are the same. 
In addition to the above properties, 
when DCs are ergodic, 
the mean FPT is infinite. 
Moreover, 
if particles take a long time to arrive at the absorbing boundary, 
the excess cumulative probability of early-arriving particles is almost zero  
and the FPT distribution can be well approximated by the L{\' e}vy-Smirnov distribution.  
We showed that these properties of diffusion with an ergodic DC result from the convergence of the time average of the DC to the ensemble average. 

There are many studies that have addressed 
properties of the FPTs for diffusion with stochastic DCs~\cite{Lanoiselee18, Sposini19, Grebenkov19, Grebenkov21, Sposini24}.  
However, 
to our knowledge, 
no studies have addressed the certainty of absorption in a general way; 
previous findings on this property remain limited to specific models~\cite{Sposini19, Grebenkov21}.  
We also believe that 
the result that 
a stochastic DC with a larger supremum leads to more efficient transport of early-arriving particles is not a previously known finding.

Our results on the efficient transport of early-arriving particles extend and generalize  
the results of previous studies~\cite{Sposini19, Sposini24}.   
The previous studies have shown that specific models including the minimal DD model achieve more efficient transport of early-arriving particles than the classical Brownian motion.  
We extended the results to non-zero stochastic DCs and generalized the results to a broad class of non-zero stochastic DCs (see also Appendix~\ref{sec:appa1}). 
However, 
the generalization of the results to cases where 
stochastic DCs can be zero, 
such as the minimal DD model, 
remains.  

The present study demonstrated that for ergodic DCs, 
the highly efficient transport of early-arriving particles almost disappears for long target distances. 
To our knowledge, 
this is the first report to identify ergodicity as a key factor in this disappearance. 
While a similar result was reported for the minimal DD model in a previous study, 
that study did not establish a link to ergodicity~\cite{Sposini24}.

%In the present study, 
%we showed that for ergodic DCs, 
%highly efficient transport of early-arriving particles almost disappears for long target distances. 
%To our knowledge, 
%no studies have shown that the ergodicity is essential for the disappearance of highly efficient transport of early-arriving particles. 
%A previous study showed a similar result for the minimal DD model, 
%but did not demonstrate a relation with the ergodicity~\cite{Sposini24}. 

%In the present study, 
%we showed that 
Our analysis revealed that 
the ergodicity of stochastic DCs is essential for 
the convergence of the FPT distribution to the L{\' e}vy-Smirnov distribution in the long-time limit.  
Although a previous study showed this convergence for the minimal DD model, 
the role of ergodicity in that convergence has not previously been established~\cite{Sposini19}.

%%In the present study, 
%%we showed that 
%Our analysis demonstrated that 
%the ergodicity of stochastic DCs is essential for 
%the convergence of the FPT distribution to the L{\' e}vy-Smirnov distribution in the long-time limit.  
%To our knowledge, 
%no studies have shown the convergence of the FPT distribution to the L{\' e}vy-Smirnov distribution from the perspective of the ergodicity of a stochastic DC. 
%In a previous study, 
%it has been already revealed that 
%the FPT distribution for the minimal DD model converges to the L{\' e}vy-Smirnov distribution in the long-time limit~\cite{Sposini19}.  
%However, 
%the concept of the ergodicity of the DC is not used to demonstrate this. 

In the present study, 
we showed that if a stochastic DC is ergodic, 
the mean FPT diverges. 
While previous studies have reported the divergence of the mean FPT in specific models where the DC can be zero~\cite{Sposini19, Lanoiselee18, Grebenkov21}, 
our result demonstrates that 
the divergence of the mean FPT holds generally for non-zero ergodic DCs. 
Specifically, 
our finding suggests that the divergence of the mean FPT observed in the minimal DD model in a previous study~\cite{Sposini19} can be reinterpreted as 
a consequence of the ergodicity of the model,  
even though our current analysis focuses on strictly positive DCs while the DC in the minimal DD model can reach zero.

In the present study, 
we revealed that 
%for diffusion in one-dimensional semi-infinite domain with an absorbing boundary, 
particles with non-zero stochastic DCs will eventually reach the absorbing boundary with probability one. 
In addition, 
we showed that when DCs are ergodic, 
the mean FPT is infinite. 
These results demonstrate that the combination of certain absorption and an infinite mean FPT
—properties also found in the classical Brownian motion—
is a robust feature that persists even in complex, 
time-varying environments, 
as long as they lead to a non-zero ergodic fluctuations in DC.
%These results indicate that for diffusion with non-zero ergodic DCs, 
%the combination of certain recurrence and an infinite mean FPT is a robust feature that persists even under complex, time-varying environments, 
%similar to classical Brownian motion. 

An infinite mean FPT is also a characteristic of the classical Brownian motion. 
While diffusion with ergodic DC exhibits a higher cumulative probability of early-arriving particles, 
this may prompt the question of whether the proportion of late-arriving particles is also higher compared to the classical case. 
However, 
Eqs.~(\ref{eq:gsDd}) and (\ref{eq:gsD}) show that the long-time approximations of the FPT distributions for both diffusion processes are identical. 
Consequently, 
there is no difference in the proportion of particles that reach the absorbing boundary at late times. 
This is consistent with the result obtained in a previous study for the minimal DD model~\cite{Sposini19}.

Eqs.~(\ref{eq:gsDd}) and (\ref{eq:gsD}) demonstrate that in the long-time limit, 
the cumulative probability of particles arrived at the boundary is comparable for both diffusion with ergodic DCs and the classical Brownian motion. 
Conversely, 
the cumulative probability of early-arriving particles is higher for diffusion with ergodic DCs. 
Combined with certain absorption of both the classical Brownian motion and diffusion with ergodic DCs, 
these findings imply the existence of time intervals during which the cumulative probability for the classical Brownian motion exceeds that 
for diffusion with ergodic DCs; 
in other words, 
a crossover in the survival probability inevitably occurs. 
While a previous study identified this crossover in the minimal DD model~\cite{Sposini19}, 
our results generalize this phenomenon, 
showing that it occurs for any ergodic fluctuation of the DC, 
provided the DC remains positive, 
although the previous study further showed that this crossover occurs universally regardless of initial conditions.

%In the present study, 
%we revealed that for diffusion in one-dimensional semi-infinite domain with an absorbing boundary, 
%particles with non-zero stochastic DCs will eventually reach the absorbing boundary with probability one. 
%In addition, 
%we showed that when DCs are ergodic, 
%the mean FPT is infinite. 
%These results indicate that for diffusion with non-zero ergodic DCs, 
%the combination of certain recurrence and an infinite mean FPT is a robust feature, 
%persisting even under complex, time-varying environments, 
%similar to the classical Brownian motion.
%%From these results, 
%%we can see that for diffusion in one-dimensional semi-infinite domain with an absorbing boundary, 
%%particles with non-zero ergodic DCs will eventually reach the absorbing boundary with probability one, 
%%but the mean FPT diverges. 
%%This property is the same as that of the classical Brownian motion. 
%%This suggests that as long as the DCs are ergodic, 
%%the recurrence property and the infinite mean FPT are robust features that persist even when the diffusion is governed by complex, time-varying environments. 
%Our findings imply that while the ergodic stochasticity in DCs introduces high transport efficiency of early-arriving particles at short times, 
%it does not alter the fundamental topological constraints of one-dimensional diffusion.

Although most of the behavior of FPTs revealed 
in the present study 
can also be observed in three-dimensional diffusion outside a spherical absorbing boundary,  
there are some important differences. 
In contrast to diffusion in a one-dimensional semi-infinite domain with an absorbing boundary, 
for three-dimensional diffusion outside a spherical absorbing boundary, 
the probability of eventual absorption is strictly less than one. 
In addition, 
for an ergodic DC, 
when the distance from the initial position of particles to the absorbing boundary increases, 
the excess cumulative probability of early-arriving particles decreases more rapidly in three-dimensional diffusion outside a spherical absorbing boundary than 
in a one-dimensional semi-infinite domain with an absorbing boundary. 
See Appendix~\ref{sec:app3} for details.

%In the present study, 
%we revealed that 
%for diffusion in one-dimensional semi-infinite domain with an absorbing boundary, 
%particles with non-zero stochastic DCs will eventually reach the absorbing boundary with probability one. 
%In addition, 
%we showed that when DCs are ergodic, 
%the mean FPT is infinite. 
%From these results, 
%we can see that 
%for diffusion in one-dimensional semi-infinite domain with an absorbing boundary, 
%particles with non-zero ergodic DCs will eventually reach the absorbing boundary with probability one, 
%but the mean FPT diverges.
%This property is the same as that of the classical Brownian motion 
%and distinguishes the diffusion in one-dimensional semi-infinite domain with an absorbing boundary from 
%the three-dimensional diffusion outside a spherical absorbing boundary.  
%%suggests that a system with a non-zero ergodic DC is an infinite-measure ergodic system. 

For the model in which a DC is described by the Cox-Ingersoll-Ross process, 
it has been shown that for short times, 
the FPT distribution can be approximated by the distribution in the superstatistical approach~\cite{Lanoiselee18, Lanoiselee24}. 
This is also true for diffusion with a broad class of stochastic DCs that are positive and non-zero.  
In a previous study, 
we showed that our approach includes the superstatistical approach as a special case: 
the DC is described by a special stochastic process in which each sample path of the DC is time independent~\cite{Uchida24}. 
When a DC is described by the special stochastic process, 
from Eq.~(\ref{eq:pdexp2gf2}), 
we have 
\begin{eqnarray}
    g_{ss}(t,x_{0}) &=& \int _{\Omega }  \frac{x_{0}}{\sqrt{4\pi D(\omega ) t^{3}}} \exp \left[ -\frac{x_{0}^{2}}{4D(\omega )t} \right] P(d\omega ) \nonumber \\
                    &=& \int_{0}^{\infty }  \frac{r_{ss}(D)x_{0}}{\sqrt{4\pi Dt^{3}}} \exp \left( -\frac{x_{0}^{2}}{4Dt} \right) dD, 
\end{eqnarray} 
where $ g_{ss}(t,x_{0}) $ represents the FPT distribution in the superstatistical approach and 
$ r_{ss}(D) $ represents the probability distribution of $ D $.  
When $ t_{s} $ is sufficiently smaller than the time that characterizes the change in $ D(t) $, 
from Eq.~(\ref{eq:pdexp2gf2}), 
for the short-time side of the FPT distribution, 
we have 
\begin{eqnarray}
    g(t,x_{0}) &\approx & \int _{\Omega }  \frac{x_{0}}{\sqrt{4\pi D(0;\omega ) t^{3}}} \exp \left[ -\frac{x_{0}^{2}}{4D(0;\omega )t} \right] P(d\omega ) \nonumber \\
                    &=& \int_{0}^{\infty }  \frac{p_{D} (D,0)x_{0}}{\sqrt{4\pi Dt^{3}}} \exp \left( -\frac{x_{0}^{2}}{4Dt} \right) dD. \label{eq:gstsst}
\end{eqnarray}
This is exactly an approximation of the short-time side of the FPT distribution by the superstatistical formula. 
Here, 
note that when the time for characterizing the change in $ D(t) $ is sufficiently longer than the observation time, 
the approximation given by Eq.~(\ref{eq:gstsst}) is essentially an approximation of the entire distribution. 

In the present study, 
we demonstrated that for ergodic DCs, 
highly efficient transport of early-arriving particles almost disappears for a distant target. 
This result indicates that a non-ergodic DC may be desirable to maintain highly efficient transport of early-arriving particles to a distant target. 
In fact, 
even after a long time, 
the tails of the propagator for displacement of diffusion with a non-ergodic DC can remain heavier than 
that of the corresponding diffusion with ensemble-averaged diffusivity~\cite{Uchida24}. 
In addition, 
in some DD models that are non-ergodic, 
the propagators for displacement remain non-Gaussian over all times~\cite{Grebenkov21, Sposini20}, 
suggesting that in these models the FPT distributions do not converge to L{\'e}vy-Smirnov distributions in the long-time limit. 

A non-Markovian DC may allow for efficient transport to more distant targets than a Markovian DC. 
Diffusion with two-state switching diffusivity has been extensively investigated~\cite{Uneyama15, Miyaguchi16, Akimoto16, Vestergaard18, Grebenkov21}. 
Miyaguchi et al. have investigated a two-state model in which the sojourn time distribution in each state is given by a power law as a model of switching diffusivity~\cite{Miyaguchi16}. 
They have shown that the propagator for displacement is non-Gaussian at short times, 
but converges to a Gaussian distribution in the long-time limit. 
They have also shown that this convergence is slow. 
This slow convergence may be coupled with the slow relaxation of the RSD of the TAMSD. 
The approximate formula for the RSD of the TAMSD given in \cite{Uneyama15, Miyaguchi16} is essentially the same as Eq.~(\ref{eq:SvarDcorrep}). 
Thus, 
for the above two-state model, 
the excess cumulative probability of early-arriving particles may not approach to zero 
unless $ t_{s} $ becomes large. 
Similarly, 
the FPT distribution may not approach to the L{\'e}vy-Smirnov distribution 
unless $ t_{s} $ becomes large. 
These suggest that 
the above two-state model may achieve efficient transport in an early arrival of particles to more distant targets than 
the two-state Markov process model with the same ensemble average of the DC.

One of the future issues is to clarify for which class of stochastic DCs that can take the value of zero the behaviors of FPTs revealed in the present study are observed. 
The properties of FPTs revealed in the present study are obtained under the assumption that stochastic DCs are non-zero. 
However, 
some of the properties are also observed for important specific stochastic DCs that can be zero. 
For example, 
previous studies have shown that specific models including the minimal DD model and a variant of the DD model achieve more efficient transport of early-arriving particles than the classical Brownian motion~\cite{Sposini19, Sposini24}. 
The divergence of the mean FPT has also been shown for the minimal DD model~\cite{Sposini19}. 

In the present study, 
we showed that diffusion with a broad class of stochastic DCs that are positive and non-zero displays more efficient transport 
of early-arriving particles than the corresponding diffusion with ensemble-averaged diffusivity. 
This result suggests that 
fluctuations in DCs may have beneficial effects on signal initiation and molecular search processes in biological, chemical, and physical systems.  
The result also suggests that fluctuations in DCs may significantly impact diffusion-limited reactions, 
especially triggered by single molecules, 
by enabling rare, 
rapid arrivals of single molecules that would not occur in systems with constant DCs.

\section{Conclusions} \label{sec:conc}
In the present study, 
we investigated how random fluctuations in DCs affect the time it takes for a diffusing particle to reach a target. 
We found that in a one-dimensional semi-infinite domain with an absorbing boundary, 
all particles are eventually absorbed, 
even when the DC is described by a positive and non-zero stochastic process. 
We also found that 
diffusion with a stochastic DC shows an efficient transport of particles early arrived at the absorbing boundary 
compared to the corresponding diffusion with ensemble-averaged diffusivity. 
In addition, 
if a stochastic DC has a larger supremum, 
diffusion shows more efficient transport of early-arriving particles, 
despite having the same ensemble average DC as other cases.  
For ergodic DCs, 
the mean FPT is infinite. 
In addition, 
in the long-time limit, 
the excess proportion of early-arriving particles becomes negligible, 
and the FPT distribution approaches the L{\' e}vy-Smirnov distribution, 
which is the FPT distribution of the classical Brownian motion. 
Although most of the results also hold in three-dimensional diffusion outside a spherical absorbing boundary, 
there are some important differences: for three-dimensional diffusion outside a spherical absorbing boundary, 
absorption is not guaranteed for all particles and for ergodic DCs, 
as the distance to the absorbing boundary increases, 
the excess proportion of early-arriving particles approaches zero more rapidly.

Our results on the efficient transport of early-arriving particles support the previous finding that 
diffusion with stochastic DCs is significantly more efficient than the classical Brownian diffusion in extreme targeting scenarios~\cite{Sposini24}. 
However, 
our results on the efficient transport for ergodic DCs suggest that 
for efficient transport of early-arriving particles to a distant target, 
it may be advantageous for fluctuations in DCs to be non-ergodic. 
Furthermore, 
a comparison between the DC described by a two-state Markov process and 
the DC described by a two-state process in which the sojourn time distribution in each state is given by a power law suggests that 
DCs described by non-Markov processes may allow for efficient transport of early-arriving particles to more distant targets. 
 
The FPT properties revealed in the present study can be observed for a broad class of stochastic DCs, 
but are restricted to non-zero DCs. 
It is a future issue 
to clarify for which class of stochastic DCs that can be zero the behaviors of FPTs revealed in the present study are observed.

\section*{Data availability}
The codes for our numerical calculations are available at GitHub~\cite{gthb}. 

\appendix
\section{Equivalence of the subordination formula with our formula of the FPT distribution}    \label{app:equivsofof}
In a one-dimensional semi-infinite domain with an absorbing boundary, 
the subordination form of the FPT distribution for diffusion with a stochastic DC is given by~\cite{Lanoiselee18,Grebenkov21}  
\begin{equation}
    g'(t,x_{0}) = \int_{0}^{\infty } \rho (t;T) g_{0} (T,x_{0}) dT,  \label{eq:soff}
\end{equation}
where $ g'(t,x_{0}) $ represents the subordination form of the FPT distribution for diffusion with a stochastic DC,   
$ \rho (t;T) $ represents the probability distribution of times when $ \int_{0}^{t} D(t') dt' $ first becomes equal to $ T $, and 
$ g_{0} (T,x_{0}) $ represents the FPT distribution for the ordinary Brownian motion with unit diffusivity and is given by
\begin{equation}
    g_{0} (T,x_{0}) = \frac{x_{0}}{\sqrt{4\pi T^{3}}} \exp \left( -\frac{x_{0}^{2}}{4T} \right) .  \label{eq:sosls} 
\end{equation}
The probability distribution of times when $ \int_{0}^{t} D(t') dt' $ first becomes equal to $ T $ 
has a relation with the probability distribution of $ \int_{0}^{t} D(t') dt' $~\cite{Lanoiselee18,Grebenkov21}   
\begin{equation}
    \int_{0}^{\infty } \exp (-\gamma T) \rho (t;T) dT = -\frac{1}{\gamma } \frac{\partial }{\partial t} \Upsilon (t;\gamma ) , \label{eq:soltr}
\end{equation}
where $ \Upsilon (t;\gamma ) $ represents the Laplace transform of the probability distribution of $ \int_{0}^{t} D(t') dt' $. 

The Laplace transform of the probability distribution of $ \int_{0}^{t} D(t') dt' $ was originally given in 
the form of the ensemble average with respect to $ \int_{0}^{t} D(t') dt' $~\cite{Grebenkov21}
\begin{equation}
    \Upsilon (t;\gamma ) = \left\langle \exp \left( -\gamma \int_{0}^{t} D\left( t' \right) dt' \right) \right\rangle . 
\end{equation}  
By giving the Laplace transform of the probability distribution of $ \int_{0}^{t} D(t') dt' $ in the form using $ P(d\omega ) $, 
we can show the equivalence of our formula to the subordination form of the FPT distribution. 
The Laplace transform of the probability distribution of $ \int_{0}^{t} D(t') dt' $ in the form using $ P(d\omega ) $ is given by  
\begin{equation}
    \Upsilon (t;\gamma ) = \int_{\Omega } \exp \left( -\gamma \int_{0}^{t} D\left( t';\omega \right) dt' \right) P(d\omega ). 
\end{equation}
From this equation, 
we have 
\begin{equation}
    -\frac{1}{\gamma } \frac{\partial }{\partial t} \Upsilon (t;\gamma ) = \int_{\Omega } \exp \left( -\gamma \int_{0}^{t} D\left( t';\omega \right) dt' \right) D(t;\omega ) P(d\omega ).
\end{equation}
Substituting this equation into Eq.~(\ref{eq:soltr}) and performing the inverse Laplace transform lead to 
\begin{equation}
    \rho (t;T) = \int_{\Omega } \delta \left( \int_{0}^{t} D\left( t';\omega \right) dt' - T \right) D(t;\omega ) P(d\omega ).  \label{eq:soq}
\end{equation}
By substituting Eqs.~(\ref{eq:sosls}) and (\ref{eq:soq}) into Eq.~(\ref{eq:soff}), 
we have  
\begin{equation}
    g'(t,x_{0}) = \int _{\Omega }  \frac{x_{0}D(t;\omega )}{\sqrt{4\pi S^{3}(t;\omega ) t^{3}}} \exp \left[ -\frac{x_{0}^{2}}{4S(t;\omega )t} \right] P(d\omega ). 
\end{equation}
Comparing this equation with Eq.~(\ref{eq:pdexp2gf2}) leads to 
\begin{equation}
    g'(t,x_{0}) = g(t,x_{0}). 
\end{equation}

As we have just shown, 
the subordination formula and our formula are equivalent.  
However, 
the viewpoints are different. 
The superensemble in the subordination formula consists of ensembles with 
a constant DC but different time flows. 
On the other hand, 
the superensemble in our formula consists of ensembles with 
the same time flow but different time evolutions of a DC. 

Both the subordination approach and our approach lead to equivalent formulas for the FPT distribution. 
Then, what is the advantage of using our approach for analysis? 
Our approach has the advantage of making it easier to solve problems 
related to the ergodicity of a stochastic DC.  
The essence of the subordination approach is time change: 
the time variable is transformed from real time to some kind of stochastic process~\cite{Chechkin17}. 
In Eq.~(\ref{eq:soff}), 
the transformed time variable is $ \int_{0}^{t} D(t') dt' $. 
Here, 
the point is that the transformed time variable is needed to be a monotonically non-decreasing variable. 
Because of this, 
in the subordination approach, 
the time average of a stochastic DC does not appear explicitly. 
On the other hand, 
in our approach, 
it is easy to explicitly use the time average of a stochastic DC.

\section{Simulations}    \label{app:sim}
We used the Euler method for numerical integration of the Langevin equation~\cite{Kloeden11}. 
\begin{equation}
    x(t+\delta t) = x(t) + \sqrt{2D(t) \delta t} \xi (t),  
\end{equation}
where $ \delta t $ represents the time step in simulations. 
The time step in simulations is 0.001 up to time one, 
0.01 from time one to time 10, and 0.1 after that. 
The number of trajectories we simulated is 20,000 to 100,000. 
Sample paths of $ D(t) $ were generated by generating an array of random numbers from Eq.~(\ref{eq:edp}).

\section{Slower decay of the left tail probability of the FPT distribution} \label{sec:appa1}
In \cite{Sposini24}, 
the authors performed an analysis focusing on rare and extreme events: an analysis using extreme mean FPT, 
which is governed by rare trajectories which are the few among the many to follow a quasi-geodesic path to the target.  
Here, 
we conduct a similar analysis focusing on the tail probability, 
which reflects the probability of rare and extreme events, 
of the FPT distribution. 
We then show that 
the left tail probability of the FPT distribution of diffusion with a stochastic DC decays more slowly than that of the corresponding diffusion with ensemble-averaged diffusivity, 
and this slower decay comes from the heavier short-time side tail of 
the FPT distribution of diffusion with the stochastic DC. 

From Eqs.~(\ref{eq:gppp}) and (\ref{eq:ppcdd}), 
when $ t $ is small enough, 
we have 
\begin{widetext}
\begin{eqnarray}
    \int_{0}^{t } g(t',x_{0} ) dt' &\approx & \int_{0}^{\infty } p_{D}(D,0) \frac{2\sqrt{Dt} \exp \left( -\frac{x_{0}^{2}}{4Dt} \right)}{\sqrt{\pi }x_{0}} dD, \label{eq:gppp0app} \\
    \int_{0}^{t} g_{d} (t',x_{0}) dt' &\approx & \frac{2\sqrt{\left\langle D(0) \right\rangle t} \exp \left( -\frac{x_{0}^{2}}{4\left\langle D(0) \right\rangle t} \right)}{\sqrt{\pi }x_{0}} . \label{eq:ppcdd0app}
\end{eqnarray}
\end{widetext}
From Eqs.~(\ref{eq:gppp0app}) and (\ref{eq:ppcdd0app}), 
we have 
\begin{equation}
    \lim_{t \to 0} \frac{\int_{0}^{t } g(t',x_{0} ) dt'}{\int_{0}^{t} g_{d} (t',x_{0}) dt'} \to \infty . \label{eq:fptdstsds}
\end{equation}
Here, 
note that $ \int_{\left\langle D(0) \right\rangle}^{\infty } p_{D} (D,0) \ne 0 $. 
From Eq.~(\ref{eq:fptdstsds}), 
we can see that 
the left tail probability of the FPT distribution of diffusion with a stochastic DC decays more slowly than that of the corresponding diffusion with ensemble-averaged diffusivity. 

The slower decay comes from the heavier short-time side tail of 
the FPT distribution of diffusion with a stochastic DC. 
From Eqs.~(\ref{eq:pdexp2gf2}) and (\ref{eq:fptdd}), 
when $ t $ is small enough, 
for the short-time side of the FPT distributions, 
we have 
\begin{eqnarray}
    g(t,x_{0}) &\approx & \int_{0}^{\infty }  \frac{p_{D} (D,0)x_{0}}{\sqrt{4\pi Dt^{3}}} \exp \left( -\frac{x_{0}^{2}}{4Dt} \right) dD, \label{eq:gstsst2} \\
    g_{d} (t,x_{0}) &\approx  & \frac{x_{0}}{\sqrt{4\pi \left\langle D(0) \right\rangle t^{3}}} \exp \left( -\frac{x_{0}^{2}}{4\left\langle D(0) \right\rangle t} \right) . \label{eq:ssapgd2}          
\end{eqnarray}
From Eqs.~(\ref{eq:gstsst2}) and (\ref{eq:ssapgd2}), 
we obtain 
\begin{equation}
    \lim_{t \to 0} \frac{g(t,x_{0} )}{g_{d} (t,x_{0})} \to \infty . \label{eq:ht2}
\end{equation}
Here, 
note that $ \int_{\left\langle D(0) \right\rangle}^{\infty } p_{D} (D,0) \ne 0 $. 
From Eq.~(\ref{eq:ht2}), 
we can see that 
the short-time side tail of the FPT distribution for diffusion with a stochastic DC is heavier than that of 
the FPT distribution for the corresponding diffusion with ensemble-averaged diffusivity. 
The result on the tail of the FPT
distribution and the result on the left tail probability indicate
that diffusion with a broad class of non-zero stochastic DCs displays more efficient transport 
of early-arriving particles than the corresponding diffusion with ensemble-averaged diffusivity. 

The result obtained in this section seems to contradict the results we obtained for an ergodic DC in the main text 
because the result obtained in this section is independent of $ t_{s} $. 
However, 
there is no contradiction. 
The result obtained in this section states that 
the left tail probability of the FPT distribution of diffusion with a stochastic DC decays more slowly than that of the corresponding diffusion with ensemble-averaged diffusivity, 
but does not state how much slower it is. 
Thus, 
even if 
the decay of 
the left tail probability of the FPT distribution of diffusion with a stochastic DC is slightly slower, 
the result derived in this section is correct. 
On the other hand, 
in the main text, 
we showed that when a DC is ergodic, 
the excess cumulative probability of early-arriving particles approaches zero as $ t_{s} \to \infty $. 
Conversely, 
this means that it does not become zero if $ t_{s} $ is finite. 
This is also true when a DC is described by the two-state Markov process that 
we used as an example. 
Even if the excess cumulative probability of early-arriving particles approaches zero rapidly after the crossover time, 
it does not become zero while $ t_{s} $ is finite.
Thus, 
there is no contradiction between the result obtained in the main text and that obtained in this section. 

It is important to note here that 
although 
the decay of 
the left tail probability of the FPT distribution for diffusion with a stochastic DC is indeed slower,  
for ergodic DCs, 
this only has practical significance when, 
for example, 
$ t_{s} $ is shorter than the crossover time.

\section{FPT distribution for three-dimensional diffusion outside a spherical absorbing boundary} \label{sec:app3}
Here, 
we derive a formula of the FPT distribution for three-dimensional diffusion outside a spherical absorbing boundary,  
reveal FPT properties, and 
clarify the similarities and differences between the properties of the FPT of three-dimensional diffusion outside a spherical absorbing boundary and 
those of diffusion in a one-dimensional semi-infinite domain with an absorbing boundary.  

A three-dimensional overdamped Langevin equation with a stochastic DC 
is given by 
\begin{equation}
    \frac{d\bm{x}(t)}{dt} = \sqrt{2D(t)} \bm{\xi }(t),  \label{eq:pifgt3d}
\end{equation}
where 
$ \bm{x}(t) $ is the three-dimensional position of the diffusing particle and 
$ \bm{x}(t) = (x_{1}(t),x_{2}(t),x_{3}(t))^{T} $. 
In Eq.~(\ref{eq:pifgt3d}), 
$ \bm{\xi }(t) $ is a vector of Gaussian white noises and 
$ \bm{\xi }(t) = \left( \xi _{1}(t), \xi _{2}(t), \xi _{3}(t) \right)^{T} $: $ \langle \bm{\xi }(t) \rangle = \bm{0} $ and 
$ \langle \xi _{i}(t) \xi _{j}(t') \rangle = \delta _{ij}\delta(t-t') \, (i,j=1,2,3) $. 
We assume that $ D(t) $ and $ \xi _{i}(t) $ are statistically independent. 

For a given sample path $ D(t;\omega ) $, 
the diffusion equation that corresponds to Eq.~(\ref{eq:pifgt3d}) is given by
\begin{equation}
    \partial _{t} G \left(\bm{x},t,\bm{x}_{0};\omega \right) = D(t;\omega ) \nabla ^{2} G \left(\bm{x},t,\bm{x}_{0};\omega \right) \; (\omega \in \Omega), \label{eq:tdfpgt3d}
\end{equation}
where $ \bm{x}_{0} = \bm{x}(0) $. 
Here, 
we assume spherical symmetry. 
When we use the spherical coordinate, 
we have the diffusion equation for radial direction:  
\begin{equation}
    \partial _{t} G \left(r,t,r_{0};\omega \right) =  D(t;\omega )\frac{1}{r^{2}} \partial _{r} \left( r^{2} \partial _{r} G \left(r,t,r_{0};\omega \right) \right) \; (\omega \in \Omega), \label{eq:tdfpgtr}
\end{equation}
where $ r_{0} = |\bm{x}_{0}| $. 
We solve this equation under the initial condition  
\begin{equation}
    G \left( r,0,r_{0};\omega \right) = \frac{1}{4\pi r_{0}^{2}} \delta (r-r_{0}),  
\end{equation}
and the natural boundary condition
\begin{equation}
    \lim_{r \to \infty }  G \left(r,t,r_{0};\omega \right) = 0,  
\end{equation} 
and the absorbing boundary condition  
\begin{equation}
    G \left(R,t,r_{0};\omega \right) = 0,    
\end{equation}
where $ R < r_{0} $. 
We then have the propagator 
\begin{widetext}
\begin{equation}
    G \left(r,t,r_{0};\omega \right) = \frac{1}{4\pi rr_{0}} \frac{1}{\sqrt{4\pi S(t;\omega ) t}} \left\{ \exp \left[ -\frac{(r-r_{0})^{2}}{4S(t;\omega )t} \right] - \exp \left[ -\frac{(r+r_{0}-2R)^{2}}{4S(t;\omega )t} \right]\right\} .    \label{eq:Gomegar}
\end{equation} 
\end{widetext}

We can obtain the FPT distribution from the equation
\begin{equation}
    g(t,r_{0};\omega ) = 4\pi R^{2} D(t;\omega ) \partial _{r} G \left(r,t,r_{0};\omega \right) \left. \right|_{r=R} ,    \label{eq:frmg}
\end{equation}
where $ g(t,r_{0};\omega ) $ represents the FPT distribution for a given sample path $ D(t;\omega ) $. 
Substituting Eq.~(\ref{eq:Gomegar}) to Eq.~(\ref{eq:frmg}) leads to 
\begin{equation}
    g(t,r_{0};\omega ) = \frac{R}{r_{0}} \frac{(r_{0} - R)D(t;\omega )}{\sqrt{4\pi S^{3}(t;\omega ) t^{3}}} \exp \left[ -\frac{(r_{0}-R)^{2}}{4S(t;\omega )t} \right] .
\end{equation}
Thus, 
we have the FPT distribution for the diffusion described by Eq.~(\ref{eq:pifgt3d}),    
\begin{equation}
    g(t,r_{0}) = \frac{R}{r_{0}} \int_{\Omega } \frac{(r_{0} - R)D(t;\omega )}{\sqrt{4\pi S^{3}(t;\omega ) t^{3}}} \exp \left[ -\frac{(r_{0}-R)^{2}}{4S(t;\omega )t} \right] P(d\omega ), \label{eq:g3dr}
\end{equation}
where $ g(t,r_{0}) $ represents the FPT distribution for the diffusion described by Eq.~(\ref{eq:pifgt3d}).  
In addition, 
the FPT distribution for the corresponding diffusion with ensemble-averaged diffusivity is given by 
\begin{equation}
    g_{d} (t,r_{0} ) = \frac{R}{r_{0}} \frac{(r_{0} - R)\left\langle D(t) \right\rangle}{\sqrt{4\pi \left\langle S(t) \right\rangle ^{3} t^{3}}} \exp \left[ -\frac{(r_{0}-R)^{2}}{4\left\langle S(t) \right\rangle t} \right] , \label{eq:fptdcdd3d}
\end{equation}
where $ g_{d} (t,r_{0} ) $ represents the FPT distribution for the corresponding diffusion with ensemble-averaged diffusivity. 
Here, 
note that as in the FPT distributions given by Eqs.~(\ref{eq:pdexp2gf2}) and (\ref{eq:fptdd}), 
the short-time side tail of the FPT distribution given by Eq.~(\ref{eq:g3dr}) is heavier than that of 
the FPT distribution given by Eq.~(\ref{eq:fptdcdd3d}). 

Unlike diffusion in a one-dimensional semi-infinite domain with an absorbing boundary,  
the probability of eventual absorption is strictly less than one. 
\begin{equation}
    \int_{0}^{\infty } g(t,r_{0}) dt = \frac{R}{r_{0}} < 1.    \label{eq:A3dtpap}
\end{equation}
The same is true for the corresponding diffusion with ensemble-averaged diffusivity.  
From Eq.~(\ref{eq:fptdcdd3d}), 
we have 
\begin{equation}
    \int_{0}^{\infty } g_{d} (t,r_{0}) dt = \frac{R}{r_{0}} .    \label{eq:A3dtpapd}
\end{equation}
From Eqs.~(\ref{eq:A3dtpap}) and (\ref{eq:A3dtpapd}), 
we can see that the total proportion of particles eventually absorbed is the same for diffusion of particles with a stochastic DC and 
for the corresponding diffusion with ensemble-averaged diffusivity. 

The excess cumulative probability of early-arriving particles is given by 
\begin{widetext}
\begin{equation}
    \int_{0}^{t'_{s}} g (t,r_{0}) - g_{d} (t,r_{0}) dt = \frac{R}{r_{0}} \int_{0}^{\infty } p(S,t'_{s}) \left[ \mathrm{erfc} \left( \frac{2}{\sqrt{S/\left\langle S(t'_{s}) \right\rangle}} \right) - \mathrm{erfc}  (2) \right] dS ,  \label{eq:3Dprg}
\end{equation}
\end{widetext}
where $ t'_{s} $ is a solution of the equation $ 4\sqrt{\left\langle S(t) \right\rangle t} = r_{0} - R $.  
Unlike diffusion in a one-dimensional semi-infinite domain with an absorbing boundary, 
for a given distance to the absorbing boundary, 
the excess cumulative probability of early-arriving particles is not determined only by the time average of 
the DC but also depends on $ r_{0} $. 

As in diffusion in a one-dimensional semi-infinite domain with an absorbing boundary, 
for $ t'_{s} $ sufficiently smaller than the time that characterizes the change in $ D(t) $, 
we can show that even if the ensemble averages of stochastic DCs are the same, 
diffusion with a stochastic DC with a larger supremum exhibits a more efficient transport in an early arrival of particles at the absorbing boundary. 
When $ t'_{s} $ is sufficiently smaller than the time that characterizes the change in $ D(t) $, 
particles begin to reach the absorbing boundary before the time change in $ D(t) $ occurs. 
We have an approximation for the FPT distribution $ g(t,r_{0}) $ on the short-time side, 
\begin{equation}
    g(t,r_{0}) \approx \frac{R}{r_{0}} \int_{0}^{\infty }  \frac{p_{D} (D,0)(r_{0}-R)}{\sqrt{4\pi Dt^{3}}} \exp \left[ -\frac{(r_{0}-R)^{2}}{4Dt} \right] dD.    \label{eq:3Dfptdstapprox} 
\end{equation}
From Eq.~(\ref{eq:3Dfptdstapprox}), 
we can see that $ g(t,r_{0}) $ on the short-time side is approximated by a superposition of L{\' e}vy-Smirnov distributions. 
Thus,    
even if the ensemble averages of stochastic DCs are the same, 
a stochastic DC with a larger supremum leads to a higher excess cumulative probability of early-arriving particles. 

Using 
$ h(t) $ and $ H(t) $, 
we can rewrite Eqs.~(\ref{eq:g3dr}) and (\ref{eq:fptdcdd3d}) as 
\begin{eqnarray}
    g(t,r_{0}) &=& \frac{R}{r_{0}} \int_{\Omega } \frac{(r_{0}-R) h(t;\omega )}{\sqrt{4\pi D_{m} H^{3}(t;\omega ) t^{3}}} \nonumber \\
                &  & \times \exp \left[ -\frac{(r_{0}-R)^{2}}{4D_{m} H(t;\omega )t} \right] P(d\omega ), \label{eq:3Drhogfgttdl} \\
    g_{d} (t,r_{0}) &=& \frac{R}{r_{0}} \frac{(r_{0}-R)}{\sqrt{4\pi D_{m} t^{3}}} \exp \left[ -\frac{(r_{0}-R)^{2}}{4D_{m}t} \right] .  \label{eq:3Dgbm2}
\end{eqnarray}

Even if we calculate the mean FPT only for particles that reach the absorbing boundary, 
as in diffusion in a one-dimensional semi-infinite domain with an absorbing boundary, 
for both diffusion with an ergodic DC and the corresponding diffusion with ensemble-averaged diffusivity, 
the mean FPT is infinite. 
For a large $ t $, 
from Eq.~(\ref{eq:3Dgbm2}), 
we have 
\begin{equation}
    g_{d} (t,r_{0}) \sim \frac{R}{r_{0}} \frac{(r_{0}-R)}{\sqrt{4\pi D_{m} t^{3}}} .   \label{eq:3DgsDd}
\end{equation} 
Thus, 
the mean FPT is infinite. 
Similarly, 
for a large $ t $, 
from Eq.~(\ref{eq:3Drhogfgttdl}), 
we have 
\begin{equation}
    g(t,r_{0}) \sim \frac{R}{r_{0}} \frac{(r_{0}-R)}{\sqrt{4\pi D_{m} t^{3}}} .   \label{eq:3DgsD}
\end{equation} 
Thus, 
the mean FPT is infinite. 

As in diffusion in a one-dimensional semi-infinite domain with an absorbing boundary, 
when $ t'_{s} \rightarrow \infty $, 
the FPT distribution for diffusion with an ergodic DC approaches the FPT distribution 
for the corresponding diffusion with ensemble-averaged diffusivity. 
From Eqs.~(\ref{eq:3Drhogfgttdl}) and (\ref{eq:limH}), 
when $ t'_{s} \rightarrow \infty $, 
we have 
\begin{equation}
    g (t,r_{0}) \sim \frac{R}{r_{0}} \frac{(r_{0}-R)}{\sqrt{4\pi D_{m} t^{3}}} \exp \left[ -\frac{(r_{0}-R)^{2}}{4D_{m}t} \right] .  \label{eq:3Dgtiapprox}
\end{equation}
From Eqs.~(\ref{eq:3Dgbm2}) and (\ref{eq:3Dgtiapprox}), 
we can see that the right-hand side of Eq.~(\ref{eq:3Dgtiapprox}) is the FPT distribution for the corresponding diffusion with ensemble-averaged diffusivity. 

For ergodic DCs,  
we have the excess cumulative probability of early-arriving particles, 
\begin{widetext}
\begin{equation}
    \int_{0}^{t'_{s}} g(t,r_{0}) - g_{d} (t,r_{0}) dt = \frac{R}{r_{0}} \int_{0}^{\infty } q(H,t'_{s} ) \left[ \mathrm{erfc} \left( \frac{2}{\sqrt{H}} \right) - \mathrm{erfc} (2) \right] dH,   \label{eq:3Dpr}
\end{equation} 
\end{widetext}
where $ t'_{s} = \frac{(r_{0}-R)^{2}}{16D_{m} } $. 

Under the condition that $ r_{0} $ is kept constant, 
from Eqs.~(\ref{eq:limH}) and (\ref{eq:3Dpr}), 
we have 
\begin{equation}
    \lim_{t'_{s} \to \infty } \int_{0}^{t'_{s}} g(t,r_{0}) - g_{d} (t,r_{0}) dt = 0.    \label{eq:3Dasympr} 
\end{equation}
Eq.~(\ref{eq:3Dasympr}) indicates that when $ t'_{s} \to \infty $ under the condition that $ r_{0} $ is kept constant, 
the excess cumulative probability of early-arriving particles approaches zero. 

When $ r_{0} \to \infty $, 
from Eqs.~(\ref{eq:limH}) and (\ref{eq:3Dpr}), 
we have 
\begin{equation}
    \lim_{r_{0} \to \infty } \int_{0}^{t'_{s}} g(t,r_{0}) - g_{d} (t,r_{0}) dt = 0.    \label{eq:3Dasymprzero} 
\end{equation}
Thus, 
when $ r_{0} \to \infty $, 
the excess cumulative probability of early-arriving particles approaches zero. 

As we have just shown, 
in both the condition where $ t'_{s} \to \infty $ while $ r_{0} $ is kept constant and 
the condition where $ r_{0} \to \infty $, 
the excess cumulative probability of early-arriving particles approaches zero. 
However, 
the reason why it approaches zero differs depending on which variable's limit is taken.  
When $ t'_{s} \to \infty $ under the condition that $ r_{0} $ is kept constant, 
the ergodicity of the DC causes the diffusion of a particle 
with a stochastic DC to approach the corresponding diffusion with ensemble-averaged diffusivity, 
resulting in the excess cumulative probability of early-arriving particles approaching zero. 
This is essentially the same as in diffusion in a one-dimensional semi-infinite domain with an absorbing boundary. 
On the other hand, 
when $ r_{0} \to \infty $, 
in addition to the effect of the ergodicity, 
the decrease in the proportion of particles that can reach the absorbing boundary causes 
the excess cumulative probability of early-arriving particles to approach zero. 

In diffusion in a one-dimensional semi-infinite domain with an absorbing boundary, 
the dependence of the excess cumulative probability of early-arriving particles on $ t_{s} $ does not change 
whether $ x_{0} $ increases or $ D_{m} $ decreases. 
On the other hand,
in three-dimensional diffusion outside a spherical absorbing boundary, 
due to the prefactor including $ r_{0} $, 
the dependence of the excess cumulative probability of early-arriving particles on $ t_{s} $ changes 
depending on whether $ r_{0} $ increases or $ D_{m} $ decreases. 
When $ r_{0} $ increases, 
the excess cumulative probability of early-arriving particles decreases more quickly.


% \bibliography{basename of .bib file}

\end{document}